\documentclass[12pt,preprint]{emulateapj}

\slugcomment{Astronomical Journal}

\shorttitle{Infrared 18 $\mu$m imaging of LIRGs}
\shortauthors{Imanishi et al.}

\begin{document}

%% LaTeX will automatically break titles if they run longer than
%% one line. However, you may use \\ to force a line break if
%% you desire.

\title{Subaru and Gemini High Spatial Resolution Infrared 18 $\mu$m
Imaging Observations of Nearby Luminous Infrared Galaxies}   

%% Use \author, \affil, and the \and command to format
%% author and affiliation information.
%% Note that \email has replaced the old \authoremail command
%% from AASTeX v4.0. You can use \email to mark an email address
%% anywhere in the paper, not just in the front matter.
%% As in the title, use \\ to force line breaks.

\author{Masatoshi Imanishi\altaffilmark{1,2}}
\affil{Subaru Telescope, 650 North A'ohoku Place, Hilo, Hawaii, 96720,
U.S.A.}
\email{masa.imanishi@nao.ac.jp}

\author{Keisuke Imase \altaffilmark{3}, Nagisa Oi \altaffilmark{3} }
\affil{Department of Astronomy, School of Science, Graduate
University for Advanced Studies (SOKENDAI), Mitaka, Tokyo 181-8588} 

\and

\author{Kohei Ichikawa}
\affil{Department of Astronomy, Kyoto University, Kyoto 606-8502, Japan} 

\altaffiltext{1}{Department of Astronomy, School of Science, Graduate
University for Advanced Studies (SOKENDAI), Mitaka, Tokyo 181-8588}

\altaffiltext{2}{National Astronomical Observatory of Japan, 2-21-1
Osawa, Mitaka, Tokyo 181-8588, Japan} 

\altaffiltext{3}{Subaru Telescope, 650 North A'ohoku Place, Hilo,
Hawaii, 96720, U.S.A.}  

\begin{abstract}

We present the results of a ground-based, high spatial resolution
infrared 18 $\mu$m imaging study of nearby luminous infrared galaxies
(LIRGs), using the Subaru 8.2-m and Gemini South 8.1-m telescopes.
The diffraction-limited images routinely achieved with these
telescopes in the $Q$-band (17--23 $\mu$m) allow us to
investigate the detailed spatial distribution of infrared emission in
these LIRGs.  
We then investigate whether the emission surface brightnesses are modest, 
as observed in starbursts, or are so high that luminous
active galactic nuclei (AGNs; high emission surface brightness energy
sources) are indicated.  
The sample consists of 18 luminous buried AGN candidates and
starburst-classified LIRGs identified in earlier infrared
spectroscopy. 
We find that the infrared 18 $\mu$m emission from the buried AGN candidates 
is generally compact, and the estimated emission surface brightnesses
are high, sometimes exceeding the maximum value observed in and
theoretically predicted for a starburst phenomenon.
The starburst-classified LIRGs usually display spatially extended
18 $\mu$m emission and the estimated emission surface brightnesses are
modest, within the range sustained by a starburst phenomenon.  
The general agreement between infrared spectroscopic and imaging energy
diagnostic methods suggests that both are useful tools for
understanding the hidden energy sources of the dusty LIRG population. 
 
\end{abstract}


\section{INTRODUCTION}

A large number of bright galaxies that emit most of their radiation in the 
infrared (8--1000 $\mu$m; $L_{\rm IR}$ $> 10^{11}L_{\odot}$), peaking at
$\sim$60 $\mu$m, were found with the {\it IRAS} infrared sky survey
\citep{soi87}.  
These are called luminous infrared galaxies (LIRGs) or ultraluminous
infrared galaxies (ULIRGs) if the infrared luminosity exceeds 
L$_{\rm IR}$ $>$ 10$^{12}$L$_{\odot}$ \citep{san88a,sam96}. 
The strong infrared emission means that powerful energy
sources are present, hidden behind dust, which absorbs most of the
primary energetic radiation; the heated dust grains re-emit this
energy as infrared thermal radiation. 
These energy sources can be starbursts (energy release by nuclear fusion 
reactions inside stars), or activity within an active galactic nucleus 
(AGN; energy conversion from gravitational energy generated by a 
mass-accreting supermassive black hole to radiative energy), 
or some combination of the two. 
Once the primary energetic radiation is converted to heat, identifying
the original energy source becomes an observational challenge. 

An effective method of investigating the energy sources of LIRGs
is to measure their emission surface brightnesses. 
In a starburst, the radiative energy generation efficiency of nuclear
fusion is only $\sim$0.5\% of Mc$^{2}$, where M is the mass of material 
used in the nuclear fusion reaction, and c is the speed of light. 
The emission surface brightness of a starburst is modest and 
cannot exceed a certain threshold ($\sim$10$^{13}$L$_{\odot}$
kpc$^{-2}$), as determined observationally \citep{wer76,meu97,soi00,rie09} 
and predicted theoretically \citep{elm99,tho05,you08}.
In contrast, the radiative energy generation efficiency of an AGN is 
much higher, at 6--42\% of Mc$^{2}$, where M is the mass of accreting
material (Bardeen 1970; Thorne 1974). 
In an AGN, high luminosity can be generated from a very compact area
around a mass-accreting supermassive black hole, producing a very high
emission surface brightness with $>$10$^{13}$L$_{\odot}$ kpc$^{-2}$. 

For the dusty LIRG population, high spatial resolution imaging
observations in the infrared are a suitable choice to constrain the
emission surface brightnesses of energy sources, because dust absorbs
most of the primary energetic radiation, infrared dust re-radiation
dominates the observed spectral energy distribution, and so traces the
intrinsic energetic radiation luminosities in LIRGs.
For a galaxy with given infrared luminosity, when the emission is
spatially unresolved, a higher spatial resolution image can constrain
the size of the emission region more strongly, and can provide a more
stringent lower limit for the emission surface brightness of an energy
source. 
Ground-based 8-m class telescopes are better suited for this
purpose than space-based infrared satellites with small apertures,
as the former provides better spatial resolution than the latter. 
To investigate the spatial distribution of infrared emission of
LIRGs from the ground, $N$-band (8--13 $\mu$m) and 
$Q$-band (17--23 $\mu$m) Earth atmospheric windows are possible choices. 
Higher sensitivity can be achieved in the $N$-band than $Q$-band, due to 
a smaller background level from Earth's atmosphere.
Thus, $N$-band observations are
more suitable for probing the overall morphology of infrared 
emission of LIRGs, including both nuclear compact and diffuse
extended components, to a fainter flux level \citep{soi01,alo06,dia08,sie08}. 
However, for the purpose of constraining the emission surface
brightnesses of energy sources in LIRG's nuclei, we believe that
$Q$-band observations are more powerful, for the following reasons. 
First, the long wavelengths of the $Q$-band are significantly less 
affected by turbulence in the Earth's atmosphere than 
shorter-wavelength bands. Thus, in the $Q$-band, diffraction-limited 
image sizes ($\sim$0$\farcs$5) are readily achievable using
ground-based 8-m class telescopes under even modestly good 
weather conditions. 
Hence, the point spread function (PSF) at $Q$ is stable over time, and
the investigation of the intrinsic spatial extent of compact emission in
LIRGs is very reliable.
Second, in the $N$-band, (1) the contribution from polycyclic aromatic
hydrocarbons (PAH) emission features, originating in star-forming activity, 
can be strong, relative to dust continuum emission, and (2) infrared
emission powered by an obscured energy source can be highly
flux-attenuated by the strong 9.7 $\mu$m silicate dust absorption feature
\citep{arm07,ima07a,ima09,vei09,ima10a}.
Suppose the case that $\sim$90\% of the total infrared (8--1000 $\mu$m)
luminosity of a LIRG is dominated by a compact, highly-obscured buried
AGN, with a small ($\sim$10\%) contribution from spatially-extended
star-forming activity. While the $N$-band emission from the buried AGN
can be highly flux-attenuated, spatially-extended PAH emission is
unattenuated. 
If we observe such a LIRG in the $N$-band, the fraction of the  
spatially-extended emission component, relative to the compact one, can
be substantially higher than the actual energetic contribution to the
total infrared luminosity, possibly misidentifying such a source as
starburst-important. This possible ambiguity is largely overcome, if we 
observe the LIRG in the $Q$-band, because (1) the contribution from PAH 
emission features is much reduced, compared to the $N$-band, and (2) 
the 18 $\mu$m silicate dust absorption feature is significantly weaker
than the 9.7 $\mu$m feature.
We thus choose the $Q$-band for our infrared imaging observations.

With the advent of high-sensitivity $Q$-band imaging cameras attached to
8-m class telescopes at good astronomical observing sites, $Q$-band
high spatial resolution imaging observations of LIRGs have now become
technically feasible. 
We have therefore performed such observations, using the Subaru 8.2-m 
telescope located on Mauna Kea, Hawaii, and the Gemini South 8.1-m telescope 
located at Cerro Pachon, Chile. 
Throughout this paper, H$_{0}$ $=$ 75 km s$^{-1}$ Mpc$^{-1}$,
$\Omega_{\rm M}$ = 0.3, and $\Omega_{\rm \Lambda}$ = 0.7 are adopted.

\section{TARGETS}

Due to the strong Earth atmospheric background emission in the $Q$-band 
(17--23 $\mu$m), sensitivity is still limited even with 8-m class
telescopes.  
Therefore, we chose nearby LIRGs that are bright in the $Q$-band as our
main targets. 
As LIRGs have highly concentrated nuclear gas and dust
\citep{sam96,soi00}, many of the putative AGNs are likely to be 
obscured along virtually all directions at the inner regions and deeply
{\it buried}. 
Such buried AGNs lack the narrow line regions at the 10--1000 pc scale,
photoionized by the AGN radiation, and so optical spectroscopy fails to
discover AGN (Seyfert) signatures. 
Our primary scientific goal is to search for such optically elusive
buried AGNs in optically non-Seyfert LIRGs, through the detection of 
high emission surface brightness energy sources. 
In fact, high spatial resolution $Q$-band imaging, as a tool 
for quantitatively estimating the emission surface brightnesses of energy
sources, is particularly effective at picking up buried AGNs with a dust
covering factor of about unity, because almost all primary energetic
radiation is converted to dust re-radiation, which can be probed with
infrared observations.
We selected LIRGs with luminous buried AGN signatures 
in X-ray, infrared, or millimeter spectra, as well as starburst-classified 
LIRGs (without buried AGN signatures) and optical Seyfert 2s (= obscured
by torus-shaped dusty medium) of interest for comparison.
A few very nearby well-studied galaxies with L$_{\rm IR}$ $<$
10$^{11}$L$_{\odot}$ were also included in our sample, because their
proximity allows us to investigate the infrared emission morphology 
in more detail on a {\it physical} scale.
Table 1 tabulates the 18 observed objects. 
The sample is heterogeneous.
Our scientific goals are (1) to strengthen any buried AGN signatures 
suggested in previously obtained data, and (2) to determine if the
emission surface brightnesses of energy sources in starburst-classified
galaxies are indeed below the maximum limit sustained by a starburst
phenomenon ($\sim$10$^{13}$L$_{\odot}$ kpc$^{-2}$), as widely argued.

\section{OBSERVATIONS AND DATA ANALYSIS}

\subsection{$Q$-band Observations}

Gemini $Q$-band (17--23 $\mu$m) imaging observations were made using
T-ReCS (Thermal-Region Camera Spectrograph; Telesco et al. 1998) on the
Gemini South 8.1-m telescope, at Cerro Pachon, Chile.
T-ReCS uses a Raytheon 320 $\times$ 240 pixel Si:As IBC array, with a 
pixel scale of 0$\farcs$09. 
The field-of-view is 28$\farcs$8 $\times $21$\farcs$6. 
The detector was read out in correlated quadruple sampling (CQS) mode
(Sako et al. 2003). 
The $Qa$ filter (18.3 $\mu$m; 17.6--19.1 $\mu$m) was
used to trace the $\sim$18 $\mu$m emission from LIRGs.
The standard chop-nod technique (telescope nodding and secondary mirror
chopping) was employed to subtract the background emission from Earth's
atmosphere and the telescope. 
Object signals were taken at four different positions: (Nod, Chop) = 
(1, A), (1, B), (2, A), and (2, B).
However, for T-ReCS data, only one chop position was guided, and 
compact, diffraction-limited-images were obtained. 
At another chop position, the telescope was unguided, and stellar images
were clearly degraded and spatially extended. 
Only the data taken at the guided chop position were used for our 
analysis, to investigate the intrinsic spatial distribution of the
infrared 18 $\mu$m emission in a reliable manner.
The chop and nod throw was 15''.
For some sources with spatially extended structures, the position angle 
of T-ReCS was arranged to simultaneously cover the emission from these 
interesting regions. 
The frame rate was 25.8 ms, and total on-source exposure time was
15--31 minutes. 
Table 2 summarizes our detailed observing log. 
Weather conditions were very good (low background level) in 2008 and
2009, but during the 2010 observing runs, a high Earth atmospheric 
background level allowed us to observe only bright sources. 

Subaru $Q$-band imaging observations were made using COMICS
\citep{kat00,oka03} on the Subaru 8.2-m telescope \citep{iye04} atop
Mauna Kea, Hawaii, under clear weather conditions. 
The details are shown in Table 2.
The precipitable water vapor value was as low as $<$1 mm during the
2008 observing run, and 3--4 mm during the 2009 run. 
The $Q17.7$ (17.7 $\mu$m; 17.25--18.15 $\mu$m) filter was used to study
the $\sim$18 $\mu$m emission from LIRGs.
COMICS employs a 320 $\times$ 240 Si:As IBC detector for imaging observations.
The pixel scale is 0$\farcs$13, providing a field of view of 42'' 
$\times$ 32'', if the full array is used.
However, as the background emission was large for $Q17.7$-band imaging
observations, we used a partial-readout mode, to avoid saturating 
the array.
Only 50--120 rows were read out, depending on weather conditions,
particularly on the precipitable water vapor value. 
The standard chop-nod technique was employed to subtract background
emission. 
The frame rate was 16--31 ms, depending on the actual background
level at the time of observations, and the total on-source exposure time
ranged from 16 to 60 minutes. 

Our Gemini T-ReCS data were reduced using the IRAF-based analysis tool provided
by the Gemini observatory 
%---
\footnote{http://www.gemini.edu/sciops/instruments/t-recs/data-format-and-reduction?q=node/10145}.
%---
After inspecting individual frames to confirm that no bad
frames were included, the task ``mireduce'' was used to combine frames
to improve the signal-to-noise ratios. 
For some frames, a stripe-pattern noise was recognizable. 
We used the task ``miclean'' to correct for this noise pattern.

The Subaru COMICS data were analyzed using the IRAF-based software package
provided by the COMICS team
%---
\footnote{
http://canadia.ir.isas.ac.jp/comics/open/rbin/rbin.html}.
%--- 
First, a flat-field image was created from frames with no object signals, and
then individual frames were divided by the flat image. 
Bad pixels and pixels hit by cosmic rays were identified, and signal
values in these pixels were interpolated from the signals of the
surrounding pixels. 
The frames were summed to improve the signal-to-noise ratios. 

For both Gemini South T-ReCS and Subaru COMICS data, flux calibrations
were made by comparing the signals of the targets with those of nearby
bright standard stars compiled by Cohen 
%---
\footnote{
http://www.gemini.edu/sciops/instruments/mir/Cohen\_list.html}.
%---
The estimated flux levels of the target objects with sufficient signal 
at different nod/chop positions agreed to within $<$0.1 mag. 
Thus, the adopted choice for the absolute flux level of standard stars 
may represent the largest uncertainty when compared with other flux
measurements reported in the literature. 

The observed nearby bright standard stars were also used to estimate the
point spread function (PSF) at the observed wavelengths, at the time
close to the target observations, in a sky direction similar 
to the targets. 
The standard star images in the $Qa$-band (18.3 $\mu$m) taken with Gemini
South T-ReCS generally showed diffraction ring patterns around bright
central peaks. 
Those in the $Q17.7$-band taken with Subaru COMICS often showed the
so-called Mitsubishi-pattern (three diamonds around central bright cores). 
Both of these observed features strongly suggest that nearly 
diffraction-limited image sizes were routinely achieved at 
$\sim$18 $\mu$m for both the Gemini South and Subaru telescopes. 

Standard stars were observed before and after the targets. 
The FWHM values differed slightly between the two datasets. 
Although the effects of seeing (Earth atmospheric turbulence) at 
$\sim$18 $\mu$m are much smaller than at shorter wavelengths, all images
can obviously suffer from a non-negligible amount of seeing-originated
PSF broadening. 
Thus, it is reasonable to assume that the smallest FWHM is the most
representative of the intrinsic PSF \citep{rad08}. 
As standard stars are bright, we can estimate the FWHM by dividing 
the combined final frame into individual frames. 
We adopted the smallest FWHM values found in some individual frames as
the PSF at $\sim$18 $\mu$m (FWHM$_{\rm PSF}$). 

The PSFs of standard stars were estimated using the IRAF task
{\it imexamine}, {\it a}.
Following \citet{rad08}, the Moffat function \citep{mof69} was basically
adopted, but direct FWHM estimates were sometimes considered as
a consistency check. 
In general, both values agreed for the data with sufficiently 
strong signals. 

For Gemini South T-ReCS data, \citet{rad08} measured the FWHM =
0$\farcs$53 $\pm$ 0$\farcs$039 at the same $Qa$-band. 
Our adopted FWHM values for the standard star (FWHM$_{\rm PSF}$) ranged 
from 0$\farcs$515 to 0$\farcs$565, consistent with the 
measurements by \citet{rad08}.
The Superantennae galaxy is at declination $-$72$^{\circ}$, and is very 
far south even for the Gemini South telescope site, resulting in a larger
air mass, even at the meridian, than other LIRGs. 
The FWHM of HR 7383, the standard star of Superantennae, was actually 
slightly larger than the FWHMs of other standard stars, possibly because
of the telescope primary mirror misalignment caused by a larger air mass. 
For Subaru COMICS data, the adopted FWHM values for the standard star
PSFs fell in the 0$\farcs$41--0$\farcs$45 range.
Thus, it is important to use the FWHM of a standard star observed
at an air mass similar to the air mass of each target.

As our target objects, LIRGs, are generally fainter than bright
standard stars, it was difficult to measure the FWHM values in
individual frames in a reliable manner, due to the low 
signal-to-noise ratios. 
Thus, the FWHM values were reliably measurable only in the combined
final frames. 
For the Gemini South T-ReCS LIRG data, one final dataset consisted of a
$\sim$15-min combined exposure, and one or two such datasets were taken.
For moderately bright LIRGs, we inspected individual frames in each
dataset, but no obvious continuous image drift was detected. 
Therefore, we measured the FWHM values in the final dataset. 
For objects with two such datasets, a smaller FWHM value was adopted. 

For Subaru COMICS data of LIRGs, one dataset consisted of 4-minute
exposures, in which we could identify object signals and roughly measure the
object peak positions.
For some LIRGs, the peak position continuously shifted along one
direction at a maximum distance of a few pixels.
Since the air mass changed as observations proceeded, this could be
explained by the differences in refraction between the optical (guiding
camera) and mid-infrared light (COMICS camera) due to the Earth's
atmosphere.  
We corrected for this shift manually at both the pixel scale (using the IRAF
task {\it imshift}) and the sub-pixel scale (using the IRAF task 
{\it imshift} and {\it imalign}).
We then produced a final combined frame, and measured the FWHM values.
The sub-pixel scale shift often provided larger FWHM values than the
pixel scale shift, possibly because of the signal split into surrounding
pixels during the sub-pixelization process. 
We adopted the smaller FWHM values as the $\sim$18 $\mu$m emission sizes
of LIRGs (FWHM$_{\rm LIRG}$). 

The intrinsic 18 $\mu$m emission sizes of LIRGs (FWHM$_{\rm intrinsic}$)
were estimated using the formula FWHM$_{\rm intrinsic}$$^{2}$ = 
FWHM$_{\rm LIRG}$$^{2}$ $-$ FWHM$_{\rm PSF}$$^{2}$. 
While the FWHM values for the standard stars were derived from individual
short exposure frames, those for LIRGs were obtained from combined frames
with longer exposure times. 
The FWHM values in the final combined frames are inevitably larger than 
the smallest FWHM values found in some individual short exposure frames, 
due to possible telescope tracking error (drift) and/or possible inclusion 
of poorer seeing data in some fraction of individual frames than the 
remaining majority of frames.
Thus, the FWHM$_{\rm intrinsic}$$^{2}$ values for LIRGs estimated in
this way are conservative upper limits, or the estimated 
emission surface brightnesses (= flux divided by size) of LIRGs 
are conservative lower limits. 
However, our main conclusions are robust to this ambiguity because the
presence of a luminous AGN is considered only if the lower limit of the
emission surface brightness of the energy source in the LIRG
substantially exceeds the maximum allowed value for a starburst
phenomenon (see $\S$5).  

Finally, as mentioned by \citet{rad08}, spectral energy distributions
differ between LIRGs and standard stars, in such a way that 
standard stars are generally bluer than LIRGs. 
Hence, the effective wavelength in the observed filters can be slightly
different (i.e., shorter for standard stars), possibly providing
different PSFs between LIRGs and standard stars. 
However, the widths of the $Qa$ (Gemini South T-ReCS) and 
$Q17.7$ (Subaru COMICS) filters used are so narrow that this effect is
practically negligible \citep{rad08}. 
Even if it were non-negligible, this effect would result in a slightly
larger PSF for LIRGs (red) than bluer standard stars, providing 
upper limits (lower limits) for the emission size (emission surface
brightnesses) of LIRGs. 
Again, this ambiguity does not alter our conclusions (see $\S$5). 

\subsection{10 $\mu$m Observations}

When precipitable water vapor values were high ($>$3 mm) during Subaru
COMICS observing runs, we performed $N$-band (8--13 $\mu$m) imaging
observations of bright sources, mostly unobscured quasars (QSOs), as
a backup program. 
The $N8.8$ (8.4--9.2 $\mu$m) or $N11.7$ (11.2--12.1 $\mu$m) filters were
used, because meaningful data could be obtained for bright sources 
in these filters (compared to the $Q17.7$ filter) even under high
precipitable water conditions. 
It is interesting to measure the emission surface brightnesses 
of these QSOs (i.e., known luminous unobscured AGNs), based
on the $\sim$10 $\mu$m observations. 
Table 3 shows our observing log.

For the 2006 observing run, only 160 rows were read out.
The standard chop-nod technique was employed to subtract the background
emission. 
The exposure times were 41--100 ms, and the total on-source exposure
time ranged from 13 to 53 minutes. 
For the 2009 observing run of the LIRG Mrk 266, observations were made
only at the nod-1 position, using the full array, to simultaneously
cover both the northern and southern nuclei with a separation of
$\sim$10''. 

The FWHMs of standard stars and target objects were derived in the same way as
outlined for the 18 $\mu$m observations.
The IRAF task {\it imexamine} was used, and the Moffat function was
basically assumed. 
For bright standard stars, data were divided into individual frames, and
the smallest FWHM value was adopted as the intrinsic PSF
because any unexpected factors (tracking errors or sudden seeing
degradation during parts of the exposure time) work to increase the FWHM
values.
As the QSOs were relatively bright, FWHM values were derivable in
individual frames (100-s exposure).
The smallest FWHM value was adopted.
As the exposure time of individual frames was longer for QSOs 
(100 s) than standard stars (5 s), this procedure still provides
conservative upper limits for the intrinsic spatial extent of 
$\sim$10 $\mu$m emission of QSOs.
The $\sim$10 $\mu$m emission of QSOs was dominated by compact components.
The LIRG Mrk 266 displayed a spatially extended emission, compared
to the adopted PSF. 

\subsection{Spitzer IRS Spectroscopy}

The presence of spatially extended, faint fuzzy emission, undetectable
with our ground-based 18 $\mu$m imaging observations, can be
investigated by comparing our data with the spectra taken by 
{\it Spitzer} IRS LL2 (14.0--21.3 $\mu$m) \citep{hou04}, if available. 
The aperture size of IRS LL2 with 10$\farcs$5 is sufficiently large to cover
any extended diffuse infrared 18 $\mu$m emission. 
As our targets are well-studied LIRGs, {\it Spitzer} IRS
LL2 spectra at 18 $\mu$m are available for many of them
\citep{bra06,rou06,arm07,ima07a,wu09}. 
However, to our knowledge, the flux-calibrated {\it Spitzer} IRS LL2 
spectra have not been published for IRAS 20551$-$4250 and VV 114 E, although 
LL2 observations have been performed according to the Spitzer observing log.
We have analyzed archival spectra of these sources.
For IRAS 15250+3609 and NGC 1377, strong 18 $\mu$m silicate dust
absorption features were found in the flux-calibrated LL2 spectra
\citep{arm07,rou06}. 
We re-analyzed the IRS spectra to estimate the
optical depth of the 18 $\mu$m silicate dust absorption feature based on
our method \citep{ima07a,ima09,ima10a}, which is required for dust
extinction correction to the AGN emission ($\S$5.1). 
For NGC 5253, Spitzer IRS SH high-resolution (R $\sim$ 600) spectrum at
18 $\mu$m, with spectral mapping mode (total covered area is
22$\farcs$2 $\times$ 33$\farcs$4) is available \citep{bei06}. 
This is used for the flux comparison at 18 $\mu$m. 
For ESO 602$-$G025, NGC 7592, and Arp 193, we find no record of
Spitzer IRS spectroscopic observations at $\sim$18 $\mu$m.

The Spitzer IRS observing log of the sources analyzed in this study
is shown in Table 4. 
All four modules, SL2 (5.2--7.7 $\mu$m), SL1 (7.4--14.5 $\mu$m), 
LL2 (14.0--21.3 $\mu$m), and LL1 (19.5--38.0 $\mu$m) were used.
We analyzed all data to obtain the full 5--35 $\mu$m low-resolution (R
$\sim$ 100) spectra. 

Data analysis was performed in a standard manner, using methods 
similar to those 
employed by \citet{ima07a}, \citet{ima09}, and \citet{ima10a}. 
The latest pipeline-processed data products available at the time of our
analysis were used. Frames taken at position A were subtracted from
those taken at position B to remove background emission, consisting mostly 
of zodiacal light. 
The spectra extracted for the A and B positions were then summed.
Wavelength and flux calibrations were made on the basis of 
the {\it Spitzer} pipeline-processed data.
For the SL1 spectra, data at $\lambda_{\rm obs}$ $>$ 14.5 $\mu$m in 
the observed frame are invalid (Infrared Spectrograph Data Handbook,
version 1.0), and were therefore removed. For the LL1 spectra, data
at $\lambda_{\rm obs}$ $>$ 35 $\mu$m were not used due to excessive
noise levels. These data are not necessary for our scientific discussion.
When flux discrepancies between SL1 and LL2 were discernible in such a
way that the LL2 flux (10$\farcs$5 wide slit) was larger than the SL1
flux (3$\farcs$7), we adjusted the SL1 (and SL2) flux to match the LL2
flux. 

\section{RESULTS}

\subsection{Imaging}

Figures 1 and 2 show the infrared 18 $\mu$m images of LIRGs obtained with
Gemini-South T-ReCS $Qa$-band (18.3 $\mu$m) and Subaru 
COMICS $Q17.7$-band (17.7 $\mu$m) observations, respectively. 
Photometric measurements are summarized in Table 5. 

Most of the bright standard stars show clear signatures of 
a diffraction ring or partial ring, strongly suggesting that
diffraction-limited images were routinely obtained at 18 $\mu$m under
the normally good, frequently achievable weather conditions at the
Gemini-South and Subaru telescopes. 
Hence, at 18 $\mu$m, the PSF is stable, and the possible spatial extents of 
LIRGs can be investigated in a reliable manner, with the data only minimally
affected by the time variation of the PSF. 
For LIRGs, only VV 114 E, ESO 602$-$G025, NGC 4945, NGC 1614, and Arp
193 display clearly detectable spatially extended structures. 
The 18 $\mu$m emission of NGC 5253 is spatially compact but resolvable. 
The remaining sources are dominated by spatially compact emission, which
is not clearly distinguishable from the PSFs of standard stars.

For VV 114 E, the two faint knots below the E-2 nucleus in Figure 1 were
not recognizable in the low spatial resolution 15 $\mu$m image taken
with {\it ISO}, but similar structures were seen at 1.1 $\mu$m
\citep{lef02}, suggesting that they are starburst knots situated between the VV
114 E and W merging nuclei. 

For NGC 4945, the elongated structures along the NE to SW direction were
also seen at shorter infrared wavelengths \citep{moo96,mar00}.

\subsection{Spectroscopy}

Figure 3 shows the final Spitzer IRS spectra.
IRAS 15250+3609, IRAS 20511$-$4250, and NGC 1377 show clearly
detectable 18 $\mu$m silicate dust absorption features, while the 
feature is weak in VV 114 E. 
The estimated optical depths, based on our method \citep{ima07a}
are $\tau_{18}$ $\sim$ 0.6, 0.7, and 0.85 for IRAS 15250+3609, 
IRAS 20511$-$4250, and NGC 1377, respectively. 

\section{DISCUSSION}

\subsection{Intrinsic Emission Surface Brightnesses of Energy Sources}

\subsubsection{Comparison of the 18 $\mu$m Flux Levels}

Our ground-based 18 $\mu$m imaging observations may not be particularly
sensitive to spatially extended diffuse emission outside the detected
compact cores. 
To test the presence of a possible fuzzy component, we compare our 
18 $\mu$m fluxes for the core components with 18 $\mu$m fluxes measured 
with {\it Spitzer} IRS spectra (10$\farcs$5 aperture or even larger for 
spectral mapping mode; $\S$3.3) in Table 5.
Given the $\sim$20\% flux uncertainties of {\it Spitzer} spectra (Spitzer
Infrared Spectrograph Data Handbook, version 1.0), 
we see significantly ($>>$20\%) smaller fluxes in our measurements only
in VV 114 E and NGC 4945, both of which clearly display
spatially extended emission structures.
Therefore, we find no strong evidence that the contribution from 
spatially extended diffuse emission outside the compact core component
is important, but is simply missed with our ground-based imaging
observations, in the majority of the observed LIRGs. 

For IRAS 08572+3915 and 15250+3609, our fluxes are slightly
($\sim$100 mJy) smaller than {\it Spitzer} IRS measurements, although the
discrepancy is close to the possible flux calibration uncertainty of
{\it Spitzer} IRS.
Even if this residual flux with $\sim$100 mJy at 18 $\mu$m were due to
spatially extended diffuse starburst emission below our detection  
limit, the expected 60 $\mu$m flux is $\sim$1 Jy, if we adopt
typical spectral energy distributions of starburst galaxies at 
17--30 $\mu$m \citep{bra06}, and {\it IRAS} 60-$\mu$m to 25-$\mu$m flux 
ratios for the prototypical starburst galaxy, M82 ($\sim$4). 
The expected 60 $\mu$m flux for the possible extended starburst
component is a factor of $>$7 smaller than the measured {\it IRAS} 
60 $\mu$m fluxes of IRAS 08572+3915 and 15250+3609 (Table 1), suggesting
that any undetected spatially extended starburst emission cannot
dominate the large infrared luminosities of these LIRGs. 

For VV 114 E, the sum of the 18 $\mu$m fluxes for the compact components
(E-1 and E-2) measured with our ground-based observations are
$>$30$\%$ lower than the {\it Spitzer} IRS measurement. 
Thus, a substantial amount of diffuse, faint emission component may be
present in VV 114 E. 

For NGC 4945, the nuclear 18 $\mu$m flux (1$\farcs$8 $\times$ 1$\farcs$8
region of the brightest point A in Figure 1) is $\sim$200 mJy.
When signals in the region of discernible emission (8$\farcs$9 $\times$
3$\farcs$6) are integrated, the 18 $\mu$m flux becomes $\sim$1000 mJy.
These fluxes are still smaller than the Spitzer IRS 18 $\mu$m flux of
$\sim$1500 mJy (10$\farcs$5 aperture), suggesting that low surface brightness 
diffuse emission may exist, and which is not detectable with our
ground-based 18 $\mu$m imaging observations in NGC 4945. 

\subsubsection{Size and Emission Surface Brightness}

The sizes of the 18 $\mu$m emission in apparent and physical scales,
as estimated by the method described in $\S$3.1, are summarized in
Table 6.
For LIRGs whose core emission is clearly extended (VV 114 E, ESO
602$-$G025, NGC 5253, NGC 1614, and Arp 193), the actual values are 
adopted for the emission size.
However, for the remaining LIRGs, image sizes are only marginally larger
than the PSFs of the corresponding standard stars. Part of this effect may
be artificially induced by the longer exposures time of LIRGs compared to
standard stars ($\S$3.1). 
For LIRGs whose 18 $\mu$m image sizes are not clearly distinguishable
from the PSFs, upper limits are shown for the sizes of the 18 $\mu$m
emission regions.  
The emission surface brightnesses are estimated by dividing the 18 $\mu$m 
continuum luminosity ($\nu$L$_{\nu}$) with the physical scales of the
18 $\mu$m emitting regions (Table 6). 

For sources for which only $N$-band imaging observations were performed 
under high precipitable water vapor conditions ($\S$3.2), we estimated  
the sizes of the 10 $\mu$m emission regions and emission surface
brightnesses from the $N$-band data, using methods similar to those
outlined above. 
Table 7 summarizes the results. 
 
\subsubsection{Flux-attenuation Correction of 18 $\mu$m
Continuum Emission} 

For LIRGs with clear 18 $\mu$m silicate dust absorption features, it
is suggested that the 17--18 $\mu$m continuum emission in the rest-frame
(outside the 18 $\mu$m silicate dust absorption feature) is obscured and
flux-attenuated by foreground absorbing dust. 
Proper correction of this flux attenuation is necessary to estimate the
intrinsic, extinction-corrected luminosities of the dust-obscured energy
sources from the 18 $\mu$m observations. 

As a dust extinction curve as long as 18 $\mu$m is not well-constrained
observationally, we combine the theoretical dust extinction curve at 
$\lambda$ $>$ 7 $\mu$m with the observed extinction curve at 
shorter wavelengths.
We adopt the following relations.

\begin{enumerate}
\item $\tau_{18}$/$\tau_{9.7}$ $\sim$ 0.3 \citep{chi06,ima07a} 
\item $\tau_{9.7}$/A$_{\rm V}$ = 0.06 \citep{roc84,roc85,rie85}
\item A$_{\rm 7\mu m}$/A$_{\rm V}$ $\sim$ 0.0225
\citep{dra84,lut96,nis08,nis09} 
\item A$_{\rm 17\mu m}$/A$_{\rm 7\mu m}$ $\sim$ 1.03 \citep{dra84}
\end{enumerate}

From these relations, we obtain A$_{\rm 17 \mu m}$ = 1.3 $\times$
$\tau_{18}$.
For example, for a source with $\tau_{18}$ = 1, the 17--18 $\mu$m continuum
flux is attenuated by a factor of $\sim$3.3. 
By adopting the $\tau_{18}$ values based on our estimate (Imanishi et
al. 2007a; $\S$ 4.2 of this paper), the corrected emission surface
brightnesses are derivable for LIRGs that exhibit clear 18 $\mu$m
silicate dust absorption features. 
The corrected values are shown in Table 6.

\subsubsection{Dust Covering Factor Correction}

Infrared observations can probe only emission absorbed by dust.
For buried AGNs embedded in dust in virtually all directions, or
starbursts whose individual stellar energy sources are surrounded by
local dust with a covering factor close to unity, the intrinsic
luminosities of primary energetic radiation can reasonably be estimated
from the infrared 18 $\mu$m observations. 
However, for AGNs with a dust covering factor significantly smaller than
unity, a large amount of energetic radiation escapes without being
absorbed by dust.
Such unabsorbed emission cannot be probed through infrared observations. 
PG QSOs are unobscured Seyfert 1 AGNs, meaning that at least the
direction along our line-of-sight is dust-free, in which case 
infrared observations can provide only lower limits for the intrinsic
emission surface brightnesses of energy sources. 
Following \citet{ris10}, we apply a factor of 5 correction 
for unobscured Seyfert 1 AGNs. 
Namely, the intrinsic emission surface brightnesses of the energy
sources are 5 times larger than those estimated from infrared
10 $\mu$m observations.
These corrected values are also summarized in Table 7.

\subsubsection{Do the 18 $\mu$m Observations Precisely Trace the Spatial
Distribution of 60 $\mu$m Emission?}

LIRGs usually show an infrared emission peak at $\sim$60 $\mu$m, and
their total infrared luminosities are dominated by emission from big dust
grains in thermal equilibrium with radiation \citep{sam96}.
In the Galactic diffuse inter-stellar medium, emission from PAHs and 
very small grains, which are transiently-heated and not in thermal
equilibrium, contributes importantly to the infrared 18 $\mu$m radiation
\citep{des90}.  
Then, a question may arise whether our 18 $\mu$m observations indeed
probe the spatial distribution of dust in thermal equilibrium heated by
the dominant energy sources in LIRGs. 
We, however, believe that the 18 $\mu$m observations are a
good method to investigate the spatial distribution of 60 $\mu$m
emission in LIRG's nuclei, for the following reasons. 

First, in the case of a normal starburst, it is very likely that PAHs,
very small grains, and big dust grains are spatially well mixed with
stellar energy sources, and are heated and/or excited by the stellar
radiation field. 
The spatial distribution of the emission from PAHs, very small
grains, and big dust grains at LIRG's cores should not change 
dramatically on a few 100 pc scale.  

Next, in the case of an AGN, PAHs are largely destroyed by strong X-rays, 
so that PAH contribution to the observed infrared radiation becomes 
negligible \citep{voi92}.
Furthermore, high radiation density around an AGN 
(1) heats big dust grains in thermal equilibrium to high temperature,
and   
(2) destroys very small grains selectively, 
both of which make the contribution from big dust grains dominant to
the infrared 18 $\mu$m radiation.  

Finally, in a buried AGN, the energy source (= a mass-accreting
supermassive blackhole) is more centrally-concentrated than the
surrounding dust, so that dust temperature is higher at the inner 
part, closer to the central energy source, than the outer part.  
Assuming approximately blackbody radiation, emission at a shorter
infrared wavelength comes from more inside than that at a longer
infrared wavelength (Figure 2 of Imanishi et al. 2007a). 
Namely, the 18 $\mu$m emission size is intrinsically smaller than 
the 60 $\mu$m emission size, around a buried AGN with
centrally-concentrated energy source geometry.
The same situation could occur, if an exceptionally
centrally-concentrated, extreme starburst is considered (Figure 1e of
Imanishi et al. 2007a).
However, even for a luminous energy source with 10$^{12}$L$_{\odot}$,
the intrinsic size difference between the 18 $\mu$m and 60 $\mu$m
emitting regions should appear on a physical scale much
smaller than 100 pc, and should not be discernible in our
spatial resolution of a few 100 pc physical scale for the majority of
the observed LIRGs.  

Summarizing, the spatial extent of the 18 $\mu$m emission is taken 
to be representative of that of the 60 $\mu$m emission at LIRG's cores 
on our physical resolution scale. 
The emission surface brightnesses estimated from our 18 $\mu$m
imaging observations can put important constraints on the nature of
energy sources at LIRG's cores. 

\subsection{Individual Sources}

In Tables 6 and 7, sources are categorized into three groups:
(1) The lower limit for the emission surface brightness of the
energy source is $\geq$10$^{13}$L$_{\odot}$ kpc$^{-2}$, the maximum
value achieved by a starburst phenomenon.
(2) The actual value is substantially smaller than the starburst upper
limit.
(3) The lower limit is smaller than the starburst upper
limit.
We discuss these results on an individual source basis, by comparison with
other data reported in the literature. 

\subsubsection{Nearby Sources with Energetically Important AGNs}

The following five sources at $z <$ 0.1 are diagnosed to be
AGN-important through infrared 2.5--35 $\mu$m spectroscopy: 
NGC 1377, IRAS 08572+3915, IRAS 15250+3609, Superantennae, and 
IRAS 20551$-$4250 (Table 1, column 10). 
The estimated emission surface brightnesses of the energy sources exceed
$\sim$10$^{13}$L$_{\odot}$ kpc$^{-2}$ (the maximum value of a starburst 
phenomenon) by a factor of $>$3 for IRAS 08572+3915 and 15250+3609
(Table 6), supporting the presence of luminous buried AGNs in these
sources. 
For Superantennae, IRAS 20551$-$4250, and NGC 1377, the lower limits of
the emission surface brightnesses only barely exceed 
$\sim$10$^{13}$L$_{\odot}$ kpc$^{-2}$.
As Superantennae is classified optically as a Seyfert 2 (Table 1), 
the dust covering factor around that AGN could be smaller than the other buried AGN
candidates (IRAS 20551$-$4250 and NGC 1377), so that the intrinsic
emission surface brightness could be higher than our estimate in Table
6. 
In these three sources, although the presence of a luminous AGN is a
possibility, the identity of the energy sources is not strongly constrained 
from our 18 $\mu$m imaging observations.

\subsubsection{Distant Sources with Energetically Important AGNs}

The five ULIRGs, IRAS 00091$-$0736, 00188$-$0856, 01004$-$2237,
01298$-$0744, and 04103$-$2838 are also classified into
buried-AGN-important galaxies through infrared 2.5--35 $\mu$m spectroscopy,
but their redshifts are larger than $z >$ 0.1 (Table 1).
The lower limits of the emission surface brightnesses are below or 
only barely larger than $\sim$10$^{13}$L$_{\odot}$ kpc$^{-2}$, so that 
the constraints on the energy sources based on our infrared imaging 
methods are weak. 
Observational constraints on the size of emission regions are
practically limited by the {\it apparent} spatial extent of the infrared
emission, which corresponds to a larger {\it physical scale} for more
distant sources. 
Higher spatial resolution imaging observations at 18 $\mu$m, using
a proposed 30-m class extremely large telescope \citep{tok10,oka10}
will be required to better constrain the energy sources of ULIRGs at 
$z >$ 0.1.  

\subsubsection{Unobscured AGNs}

In Table 7, the emission surface brightnesses of five PG QSOs (=
unobscured Seyfert 1 AGNs) are estimated from 10 $\mu$m imaging data,
and their lower limits are close to or barely exceed
$\sim$10$^{13}$L$_{\odot}$ kpc$^{-2}$. 
Although these results are consistent with the AGN-dominated picture of
these PG QSOs, further quantitative discussions are difficult because of
(1) the uncertainty of the dust covering factor correction, and (2) the 
possibly lower PSF stability of the 10 $\mu$m imaging data compared to
our 18 $\mu$m results ($\S$1). 

\subsubsection{Starburst and AGN Co-existing Sources}

In four LIRGs, NGC 2623, NGC 7592 W, VV 114 E, and Mrk 266 N, the
presence of an obscured AGN is suggested, but starburst activity is also
strong (Table 1, column 10).
The estimated lower limits for the emission surface brightnesses are 
far below $\sim$10$^{13}$L$_{\odot}$ kpc$^{-2}$, and so our data do not
provide evidence for energetically important AGNs. 
 
NGC 4945 also belongs to this class. Due to its proximity 
(distance $\sim$ 4.0 Mpc), the nuclear region is spatially resolved at
small physical scales (1'' corresponds to $\sim$20 pc).
The bright region, denoted ``A'' in Figure 1, is observed as an elongated 
peak in the infrared $L$-band (3.5 $\mu$m) and $N$-band (10 $\mu$m) 
\citep{moo96,mar00}, but at 18 $\mu$m, it is barely resolved into two
distinct peaks with a separation of $\sim$0$\farcs$5, possibly
because of reduced dust extinction at 18 $\mu$m. 
The left (NE) peak is even brighter; this might be the location of the
putative buried AGN suggested from X-ray observations at $>$10 keV 
\citep{iwa93,don96,gua00}.
Two fainter distinct emission components are seen at the SW
of nucleus A.
A similar pattern is seen in the high spatial resolution infrared
$K$-band (2.2 $\mu$m) image \citep{mar00}.
This is likely due to starburst knots. 

\subsubsection{Starburst-dominant Sources}

Three LIRGs, NGC 1614, Arp 193, and ESO 602$-$G025, have no obvious
luminous AGN signatures in any observational data obtained to date, and
are consistent with a starburst-dominant picture.
These LIRGs were observed as a control sample to determine whether these
starburst-classified galaxies show spatially extended structures and 
emission surface brightnesses within the starburst range. 
As expected, spatially extended emission structures and low emission surface
brightnesses are confirmed. 
Thus, for these LIRGs, our 18 $\mu$m imaging data support a 
starburst-dominant nature. 

NGC 5253 is a very nearby galaxy (distance $\sim$ 3.9
Mpc), dominated by a nuclear super star cluster \citep{gor01,van04,alo04}. 
We observed this galaxy to determine whether emission surface brightness can
substantially exceed the maximum value for a starburst phenomenon 
($\sim$10$^{13}$L$_{\odot}$ kpc$^{-2}$), if only a super star cluster
region is extracted. 
The 18 $\mu$m emission is spatially compact, but is resolved.
The emission surface brightness estimated based on our 18 $\mu$m image 
is 1.5 $\times$ 10$^{13}$L$_{\odot}$ kpc$^{-2}$, roughly comparable 
to the maximum allowed value by a starburst phenomenon.

\subsection{Energy Sources}

A natural explanation for the energy sources with emission surface
brightnesses considerably higher than $\sim$10$^{13}$L$_{\odot}$
kpc$^{-2}$ is an AGN, because (1) AGN activity can have high radiative
energy generation efficiency ($\S$1) and (2) even a super star cluster 
produces only $\sim$10$^{13}$L$_{\odot}$ kpc$^{-2}$ ($\S$5.2.5).
In fact, high emission surface brightnesses are preferentially indicated
for LIRGs with luminous AGN signatures in other observational data, such
as infrared 2.5--35 $\mu$m spectroscopy and/or X-ray observations at E $>$ 2
keV (Table 1, column 10). 

\citet{tho09} argued that the emission surface brightness of a starburst
phenomenon can exceed $\sim$10$^{13}$L$_{\odot}$ kpc$^{-2}$, if the average
dust temperature is T $>$ 200K.
Theoretically, the maximum surface brightness of a starburst phenomenon is
determined by the balance of stellar radiation pressure to dust and
gravitational boundaries \citep{tho05}. 
Under high stellar radiation density, dust and gas
(collisionally coupled with dust) are expelled away, and further
star formation activity is suppressed \citep{tho05}. 
At a high average dust temperature, T $>$ 200K, some fraction of
small, fragile dust can be sublimated, decreasing the total dust opacity
to stellar radiation, and possibly increasing the maximum allowed
emission surface brightness of a starburst. 
If this were the case in many LIRGs, the constraint of emission
surface brightnesses of $>>$10$^{13}$L$_{\odot}$ kpc$^{-2}$ would not
serve as a strong AGN signature. 

Observationally, the spectral energy distributions of local LIRGs
usually show peaks at $\sim$60 $\mu$m, indicating typical
dust temperatures of T $\sim$ 50 K \citep{cle10}. 
For dust temperatures of T $>$ 200 K, the infrared emission should
peak at $\sim$15 $\mu$m, but no such sources are found in our
sample (Table 1). 
\citet{rie09} also calculated that the maximum dust temperature for a 
starburst is T $\sim$ 88K.
At present, the proposed starburst with T $>$ 200 K is not strongly 
supported from the observational point of view. 

The only way to reconcile the T $>$ 200 K starburst scenario and the
observed infrared spectral energy distribution in LIRGs is that the 
T $>$ 200 K starbursts are occurring only in the centrally concentrated
regions of LIRG's nuclei, and are surrounded by large amounts of nuclear
dust (similar to buried AGN geometry; Figure 1e of Imanishi et al. 2007a). 
In this geometry, the original T $>$ 200 K starburst emission is once
absorbed by the surrounding nuclear dust, and the re-emitted infrared
dust emission indicates cooler temperatures, as observed in actual
LIRGs. 
However, LIRG's nuclear regions are known to contain large amounts of 
high-density molecular gas, sufficient to trigger star formation
\citep{gao04,ima07b,wil08,ima09b}. 
Thus, it seems reasonable for star formation to occur in the large 
volumes of LIRG's nuclear regions occupied by high-density gas, and it
is not clear why stars are formed only in the centrally concentrated
regions of LIRG's nuclei. 
Therefore, we consider that high emission surface brightness energy sources 
at LIRG's cores are good signatures of AGNs, if not definitive AGN evidence. 

\section{SUMMARY}

The results of ground-based, high spatial resolution infrared 18 $\mu$m
imaging observations of nearby LIRGs were presented. 
The intrinsic size of the infrared 18 $\mu$m emitting regions was
estimated by comparison with the PSFs of standard stars, using (close to)
diffraction-limited images. Emission surface brightnesses were then 
constrained.
We drew the following conclusions from our observations: 

\begin{enumerate}
\item LIRGs with luminous AGN signatures seen in previous infrared
spectroscopy and/or X-ray observations at $>$2 keV tended to 
show compact 18 $\mu$m emission, while starburst-classified
LIRGs with no obvious AGN signatures in previous observations 
displayed spatially extended morphologies with low emission surface
brightnesses. 
For some fraction of the former LIRGs at $z <$ 0.1, the lower limits of
the emission surface brightnesses were constrained to
$>>$10$^{13}$L$_{\odot}$ kpc$^{-2}$, higher than the maximum value 
sustained by a starburst phenomenon, as argued theoretically and 
actually observed.
A luminous AGN is a natural energy source that can produce such a high
emission surface brightness. 

\item For distant LIRGs at $z >$0.1, classified as AGN important 
through previous observations, 18 $\mu$m emission was generally
spatially compact. 
However, the lower limits of the emission surface brightness were below 
or only barely exceed the upper limit expected from a starburst phenomenon. 
We could not place strong constraints on the presence of luminous AGNs for
these distant sources, simply because the spatial resolution of
ground-based 8-m class telescopes is still insufficient at 18 $\mu$m. 

\item We estimated the emission surface brightness of the super star
cluster in the very nearby galaxy NGC 5253 to be
$\sim$10$^{13}$L$_{\odot}$ kpc$^{-2}$. This did not support the
possibility that 
the emission surface brightness could be $>>$10$^{13}$L$_{\odot}$
kpc$^{-2}$ if only a super star cluster region is extracted. 

\end{enumerate}

\acknowledgments

We thank T. Fujiyoshi, A. Hatakeyama, S. Harasawa, C. H. Peng, and
K. Matsui for their support during our Subaru COMICS observing runs, and
J. Radomski, A. Lopez, M. Edwards, E. Christensen, and L. Fuhrman for
their support during our Gemini South T-ReCS observing runs.   
We are grateful to the anonymous referee for his/her very valuable
comments, and C. Packham for his advice about T-ReCS data analysis.
M.I. was supported by Grants-in-Aid for Scientific Research (no. 19740109,
22012006).  
This work is based data collected at Subaru Telescope, which is operated
by the National Astronomical Observatory of Japan, and on observations
obtained at the Gemini Observatory, which is operated by the Association
of Universities for Research in Astronomy, Inc., under a cooperative
agreement with the NSF on behalf of the Gemini partnership: the National
Science Foundation (United States), the Science and Technology
Facilities Council (United Kingdom), the National Research Council
(Canada), CONICYT (Chile), the Australian Research Council (Australia),
Ministerio da Ciencia e Tecnologia (Brazil) and Ministerio de Ciencia,
Tecnologia e Innovacion Productiva (Argentina). 
This research made use of the SIMBAD database, operated at CDS,
Strasbourg, France, and the NASA/IPAC Extragalactic Database (NED),
which is operated by the Jet Propulsion Laboratory, California Institute
of Technology, under contract with NASA. 

%\clearpage

%%%%%%%%%% Table 1 %%%%%%%%%
\begin{deluxetable}{lcrrrrcrcc}
\tabletypesize{\scriptsize}
%\rotate
\tablecaption{Observed LIRGs and their {\it IRAS}-based
infrared emission properties
\label{tbl-1}}
\tablewidth{0pt}
\tablehead{
\colhead{Object} & \colhead{Redshift}   & 
\colhead{f$_{\rm 12}$}   & 
\colhead{f$_{\rm 25}$}   & 
\colhead{f$_{\rm 60}$}   & 
\colhead{f$_{\rm 100}$}  & 
\colhead{log L$_{\rm IR}$} & 
\colhead{f$_{25}$/f$_{60}$} & 
\colhead{Optical}  & \colhead{Buried} \\
\colhead{} & \colhead{}   & \colhead{(Jy)} & \colhead{(Jy)} 
& \colhead{(Jy)} & \colhead{(Jy)}  & \colhead{L$_{\odot}$} & \colhead{}
& \colhead{Class} & \colhead{AGN}  \\
\colhead{(1)} & \colhead{(2)} & \colhead{(3)} & \colhead{(4)} & 
\colhead{(5)} & \colhead{(6)} & \colhead{(7)} & \colhead{(8)} & 
\colhead{(9)} & \colhead{(10)} 
}
\startdata
IRAS 00091$-$0738 & 0.118 & $<$0.07 & 0.22 & 2.63 & 2.52 & 12.2 & 0.08
(C) & HII & a,b,c \\  
IRAS 00188$-$0856   & 0.128 & $<$0.12 & 0.37 & 2.59 & 3.40 & 12.3 & 0.14
(C) & LI & a,b,c,d \\  
IRAS 01004$-$2237 & 0.118 & 0.11 & 0.66 & 2.29  & 1.79  & 12.3 & 0.29
(W) & HII & a,b,c \\ 
IRAS 01298$-$0744 & 0.136 & $<$0.12 & 0.19 & 2.47 & 2.08 & 12.3 & 0.08
(C) & HII & a,b,c \\  
IRAS 04103$-$2838 & 0.118 & 0.08 & 0.54 & 1.82  & 1.71  & 12.2 & 0.30
(W) & LI & a,b,e \\  
IRAS 08572$+$3915   & 0.058 & 0.32 & 1.70 & 7.43  & 4.59  & 12.1 & 0.23
(W) & LI & a,b,c,d,f,g,h,i,j \\  
IRAS 15250$+$3609   & 0.055 & 0.16 & 1.31 & 7.10 & 5.93 & 12.0 & 0.18
(C)& LI & b,c,h,k \\  
Superantennae (IRAS 19254$-$7924) & 0.062 & 0.22 & 1.24 & 5.48  & 5.79
& 12.1 & 0.23 (W) & Sy2 & j,l,m,n \\  
IRAS 20551$-$4250 & 0.043 & 0.28 & 1.91 & 12.78 & 9.95  & 12.0 & 0.15
(C) & LI (HII) & c,k,o,p,q \\  
NGC 1377          & 0.006 & 0.56 & 1.93 & 7.43 & 5.95  & 10.1 & 0.26
(W) & Unc & r,s,t \\
NGC 1614          & 0.016 & 1.38 & 7.50 & 32.12 & 34.32 & 11.6 & 0.23 
(W) & HII & \nodata \\  
NGC 2623          & 0.018 & 0.21 & 1.81 & 23.74 & 25.88 & 11.5 & 0.08
(C) & Unc & u,v \\  
NGC 7592 & 0.024 & 0.26 & 0.97 & 8.05 & 10.58 & 11.3 & 0.12 (C) & Sy2
(W),HII (E) & k \\
Arp 193           & 0.023 & 0.25 & 1.42 & 17.04 & 24.38 & 11.6 & 0.08
(C)& LI & \nodata \\  
ESO 602$-$G025 (IRAS 22287$-$1917) & 0.025 & 0.27 & 0.91 & 5.42 & 9.64 & 11.3  &
0.17 (C) & LI & \nodata \\  
VV 114 E           & 0.020 & 1.03 \tablenotemark{A} & 3.65
\tablenotemark{A} & 22.93 \tablenotemark{A} & 31.55 \tablenotemark{A} &
11.7 \tablenotemark{A} & 0.16 (C) \tablenotemark{A} & HII & i,w \\
NGC 4945 & 3.9 Mpc \tablenotemark{B} & 3.95 & 14.45 & 359.3 & 620.5 &
10.2 & 0.04 (C) & HII/LI & x,y,z,aa \\  
NGC 5253 & 4.0 Mpc \tablenotemark{C} & 2.61 & 11.96 & 30.51 & 29.36 & 9.3 & 0.39 (W) &
HII (WR) & \nodata \\  
\hline
\enddata

\tablenotetext{A}{Emission from the eastern (VV 114 E) and western 
(VV 114 W) nuclei combined.} 
\tablenotetext{B}{Adopted from \citet{ber92} and \citet{mau96}.} 
\tablenotetext{C}{Adopted from \citet{mau96} and \citet{thi03}.}

\tablecomments{
Col.(1): Object name. 
ULIRGs are listed first, and then LIRGs are placed.
The two very nearby galaxies, NGC 4945 and 5253 are put last. 
Col.(2): Redshift. For the two very nearby sources, NGC 4945 and 5253, 
distances are shown in Mpc.
Col.(3)--(6): f$_{12}$, f$_{25}$, f$_{60}$, and f$_{100}$ are 
{\it IRAS} fluxes at 12 $\mu$m, 25 $\mu$m, 60 $\mu$m, and 100 $\mu$m,
respectively, taken from \citet{kim98}, \citet{san03}, or IRAS {\it FSC}
catalog. 
Col.(7): Decimal logarithm of infrared (8$-$1000 $\mu$m) luminosity
in units of solar luminosity (L$_{\odot}$), calculated with
$L_{\rm IR} = 2.1 \times 10^{39} \times$ D(Mpc)$^{2}$
$\times$ (13.48 $\times$ $f_{12}$ + 5.16 $\times$ $f_{25}$ +
$2.58 \times f_{60} + f_{100}$) ergs s$^{-1}$ \citep{sam96}.
Col.(8): {\it IRAS} 25-$\mu$m to 60-$\mu$m flux ratio.
LIRGs with f$_{25}$/f$_{60}$ $<$ 0.2 and $>$ 0.2 are
classified as cool and warm sources (denoted as ``C'' and ``W''),
respectively, \citep{san88b}.
Col.(9): Optical spectral classification. 
``HII'', ``LI'', ``Sy2'', and ``Unc'' mean HII-region, LINER, Seyfert 
2, and optically unclassified, respectively. 
For most sources, the classification is based on \citet{vei95} and
\citet{vei99}. 
For Superantennae and IRAS 20551$-$4250, the classification by
\citet{mir91} and \citet{duc97}, respectively, is adopted.
For NGC 4945, we adopt the classification by \citet{hec90} and
\citet{moo96}. 
For NGC 5253, the classification is from \citet{con91} and
\citet{wal87}, and ``WR'' means a Wolf Rayet galaxy. 
For NGC 7592, ``W'' and ``E'' indicate the western and eastern nuclei,
respectively. 
Col.(10): Buried (obscured) AGN signatures and references. 
(a): \citet{ima07a}. (b) \citet{vei09}. (c): \citet{nar10}. 
(d): \citet{idm06}. (e): \citet{ten08}.
(f): \citet{dud97}. (g): \citet{imd00}. (h): \citet{arm07}. 
(i): \citet{ima07b}. (j): \citet{ima08}. (k): \citet{ima10b}. 
(l): \citet{ris03}. (m): \citet{bra03}. (n): \citet{bra09}.
(o): \citet{fra03}. (p): \citet{ris06}. (q): \citet{san08}.
(r): \citet{rou06}. (s): \citet{ima06}. (t): \citet{ima09b}. 
(u): \citet{mai03}. (v): \citet{eva08}. (w): \citet{lef02}. 
(x): \citet{iwa93}. (y): \citet{don96}. (z): \citet{gua00}. 
(aa): \citet{ito08}. 
}

\end{deluxetable}

%\clearpage

%%%%%%%%%% Table 2 %%%%%%%%%
\begin{deluxetable}{llcc}
\tabletypesize{\small}
\tablecaption{Log for 18 $\mu$m observations \label{tbl-2}}
\tablewidth{0pt}
\tablehead{
\colhead{Object} & \colhead{Date} & \colhead{Integration time [min]} & 
\colhead{Telescope/instrument} \\ 
\colhead{(1)} & \colhead{(2)} & \colhead{(3)} & \colhead{(4)} 
}
\startdata 
Superantennae (1)    & 2008 Sep 10 & 15.5 & Gemini-S/T-ReCS \\
IRAS 20551$-$4250 & 2008 Sep 10 & 15.5 & Gemini-S/T-ReCS \\
IRAS 01004$-$2237 & 2008 Sep 10 & 31   & Gemini-S/T-ReCS \\
IRAS 04103$-$2838 & 2008 Sep 10 & 31   & Gemini-S/T-ReCS \\ 
VV 114 E          & 2008 Sep 10 & 15.5 & Gemini-S/T-ReCS \\
NGC 1377          & 2008 Sep 10 & 15.5 & Gemini-S/T-ReCS \\
Superantennae (2) & 2009 Sep 8  & 15.5 & Gemini-S/T-ReCS \\ 
IRAS 00091$-$0738 & 2009 Sep 8  & 31   & Gemini-S/T-ReCS \\ 
IRAS 00188$-$0856 & 2009 Sep 8  & 31   & Gemini-S/T-ReCS \\ 
IRAS 01298$-$0744 & 2009 Sep 8  & 31   & Gemini-S/T-ReCS \\ 
ESO 602$-$G025    & 2009 Sep 8  & 15.5 & Gemini-S/T-ReCS \\ 
NGC 4945          & 2010 May 31 & 15.5 & Gemini-S/T-ReCS \\
NGC 5253          & 2010 May 31 & 15.5 & Gemini-S/T-ReCS \\
NGC 7592          & 2010 Aug 30 & 15.5 & Gemini-S/T-ReCS \\
 \hline
IRAS 08572$+$3915 (1) & 2008 Jan 24 & 32   & Subaru/COMICS \\  
NGC 1614          & 2008 Jan 24 & 16   & Subaru/COMICS \\
NGC 2623          & 2008 Jan 24 & 40   & Subaru/COMICS \\
IRAS 08572$+$3915 (2) & 2009 Apr 7  & 32   & Subaru/COMICS \\  
IRAS 15250$+$3609 & 2009 Apr 7  & 60   & Subaru/COMICS \\  
Arp 193           & 2009 Apr 7  & 56   & Subaru/COMICS \\  \hline
\enddata

\tablecomments{
Col.(1): Object name.
The observed sources are sorted based on the telescope used and
observing date. 
Sources above and below the horizontal solid line were observed with
Gemini-South and Subaru, respectively.
Superantennae and IRAS 08572+3915 were observed twice. 
Col.(2): Observing date in UT. 
Col.(3): Net on-source integration time in minutes.
Col.(4): Telescope and instrument.
}

\end{deluxetable}

%\clearpage

%%%%%%%%%% Table 3 %%%%%%%%%
\begin{deluxetable}{llcccc}
%\tabletypesize{\scriptsize}
\tabletypesize{\small}
\tablecaption{Log for $N$-band (8--13 $\mu$m) observations, using
Subaru COMICS \label{tbl-3}}
\tablewidth{0pt}
\tablehead{
\colhead{Object} & \colhead{Redshift} & \colhead{Type} & \colhead{Date}
& \colhead{Filter} & \colhead{Integration}  \\ 
\colhead{} & \colhead{} & \colhead{} & \colhead{} & \colhead{} &
\colhead{Time [min]} \\  
\colhead{(1)} & \colhead{(2)} & \colhead{(3)} & \colhead{(4)} &
\colhead{(5)} & \colhead{(6)}   
}
\startdata 
I Zw 1 (PG 0050$+$124) & 0.061 & Sy1 $^{a}$ & 2006 Oct 4 & N11.7 & 26.7 \\  
Mrk 1014 (PG 0157$+$001) & 0.164 & Sy1 $^{a,b}$ & 2006 Oct 4 & N11.7 & 26.7 \\  
PG 0844$+$349 & 0.064 & Sy1 $^{a}$ & 2006 Oct 4 & N11.7 & 53.3 \\
PG 2130$+$099 & 0.061 & Sy1 $^{a}$ & 2006 Oct 4 & N11.7 & 13.3 \\
PG 2214$+$139 & 0.067 & Sy1 $^{a}$ & 2006 Oct 4 & N11.7 & 33.3 \\
Mrk 266 (NGC 5256) & 0.028 & Sy2 (N) + HII (S) $^{b} $& 2009 Apr 7 & N8.8 & 13 \\
\enddata

\tablecomments{
Col.(1): Object name.
Col.(2): Redshift.
Col.(3): Optical spectral classification. 
$^{a}$: \citet{sch83}.
$^{b}$: \citet{vei99}.
For the LIRG Mrk 266, ``N'' and ``S'' denote the northern and southern
nuclei, respectively. 
Col.(4): Observing date in UT. 
Col.(5): Filter used for the $N$-band observation.
$N8.8$ (8.4--9.2 $\mu$m) or $N11.7$ (11.2--12.1 $\mu$m).
Col.(6): Net on-source integration time in minutes.
}

\end{deluxetable}

%%%%%%%%%% Table 4 %%%%%%%%%

\begin{deluxetable}{lrlcccc}
%\tabletypesize{\scriptsize}
\tabletypesize{\small}
\tablecaption{{\it Spitzer} IRS observing log
\label{tbl-2}}
\tablewidth{0pt}
\tablehead{
\colhead{Object} & \colhead{PID} &\colhead{Date} & \multicolumn{4}{c}
{Integration time [sec]} \\ 
\colhead{} & \colhead{} & \colhead{[UT]} & \colhead{SL2} & \colhead{SL1} &
\colhead{LL2} & \colhead{LL1} \\ 
\colhead{(1)} & \colhead{(2)} & \colhead{(3)} & \colhead{(4)} & 
\colhead{(5)} & \colhead{(6)} & \colhead{(7)} 
}
\startdata 
IRAS 15250$+$3609 & 105 & 2004 Mar 4  & 84 & 84 & 120 & 120 \\
IRAS 20551$-$4250 & 105 & 2004 May 14 & 56 & 56 & 56 & 56 \\
NGC 1377          & 159 & 2005 Feb 18 & 252 \tablenotemark{a} & 252 \tablenotemark{a} & 600 \tablenotemark{a} & 600 \tablenotemark{a} \\
VV 114 E  & 21  & 2004 Jun 28 & 210 \tablenotemark{b} & 210 \tablenotemark{b} & 180 \tablenotemark{b} & 180 \tablenotemark{b} \\
\enddata

\tablenotetext{a}{IRS spectral mapping mode was employed.} 

\tablenotetext{b}{IRS spectral mapping mode was employed. 
Total exposure time covering the E-nucleus was extracted.} 

\tablecomments{
Col.(1): Object name.
Col.(2): PID number: 105 (PI = J. Houck), 159 (PI = R. Kennicutt), and 
21 (PI = J. Houck).
Col.(3): Observing date in UT. 
Col.(4)--(7): Net on-source integration time for SL2, SL1, LL2, and LL1 
spectroscopy, respectively, in seconds.
}

\end{deluxetable}

%%%%%%%%%% Table 5 %%%%%%%%%
\begin{deluxetable}{lrclr}
\tabletypesize{\small}
\tablecaption{Flux measurement of LIRGs at $\sim$18 $\mu$m \label{tbl-5}}
\tablewidth{0pt}
\tablehead{
\colhead{Object} & \colhead{Flux (This work)} & \colhead{Flux (Spitzer IRS)} &
\colhead{Standard star} & \colhead{Std flux} \\ 
\colhead{} & \colhead{(mJy)} & \colhead{} & \colhead{} & \colhead{(Jy)} \\ 
\colhead{(1)} & \colhead{(2)} & \colhead{(3)} & \colhead{(4)} &
\colhead{(5)}  }
\startdata 
Superantennae (1) & 595  & $\sim$600 & HR 7383 & 2.4 \\
IRAS 20551$-$4250 & 485  & $\sim$470 & HR 7869 & 6.1 \\
IRAS 01004$-$2237 & 330  & $\sim$270 & HR 400 & 2.9 \\
IRAS 04103$-$2838 & 290  & $\sim$250 & HR 1231 & 34.6 \\ 
VV 114 E-1        & 245  & $\sim$1000 (452$\pm$61 \tablenotemark{a}) & HR 334 & 6.1 \\
VV 114 E-2        & 390  & $\sim$1000 (452$\pm$61 \tablenotemark{a}) & HR 334 & 6.1 \\
NGC 1377          & 590  & $\sim$500 & HR 1231 & 34.6 \\
Superantennae (2) & 640  & $\sim$600 & HR 7383 & 2.4 \\ 
IRAS 00091$-$0738 & 50   & $\sim$50 & HR 117 & 2.6 \\ 
IRAS 00188$-$0856 & 85   & $\sim$70 & HR 117 & 2.6 \\ 
IRAS 01298$-$0744 & 55   & $\sim$45 & HR 334 & 6.1 \\ 
ESO 602$-$G025     & 155  & \nodata & HR 8015 & 2.0 \\ 
NGC 5253          & 5400 & $\sim$6000 (5500$\pm$539 \tablenotemark{b}) & HR 5287
& 7.3 \\ 
NGC 7592 W \tablenotemark{c} & 210  & \nodata & HR 8841 & 2.6 \\
NGC 4945 A (1.8'' $\times$ 1.8'') & 200  & $\sim$1500 & HR 4906 & 4.8 \\
NGC 4945 (3.6'' $\times$ 8.9'') & 1000 & $\sim$1500 & HR 4906 & 4.8 \\ 
\hline
IRAS 08572$+$3915 (1) & 420  & $\sim$490 (552$\pm$46 \tablenotemark{d}) & HR 3275 & 8.7 \\  
NGC 1614          & 2970 & $\sim$2000 & HR 1231 & 36.8 \\
NGC 2623          & 260  & $\sim$300 & HR 3275 & 8.7 \\
IRAS 08572$+$3915 (2) & 390 & $\sim$490 (552$\pm$46 \tablenotemark{d}) & HR 3705 & 25.1 \\  
IRAS 15250$+$3609   & 210  & $\sim$300 & HR 5763 & 4.7 \\  
Arp 193           & 275   & \nodata & HR 5219 & 13.4 \\ \hline 
\enddata

\tablenotetext{a}{Measurement at 17.9 $\mu$m by \citet{soi01} for 
VV 114 E.} 

\tablenotetext{b}{Measurement at 17.4 $\mu$m (5500$\pm$539 mJy; 
Moorwood \& Glass 1982).
A small 18.7 $\mu$m flux measurement (2900 mJy) by \citet{gor01} was
commented by \citet{van04}.} 
  
\tablenotetext{c}{Both the western (W) and eastern (E) nuclei were
observed simultaneously, but the E nucleus was too faint to allow
meaningful discussion.}

\tablenotetext{d}{Measurement by Soifer et al.(2000) at 17.7 $\mu$m,
using IRAS flux for a standard star.}

\tablecomments{
Col.(1): Object name. 
The horizontal solid line separates sources observed with Gemini-South 
(above the line) and Subaru (below the line).
Superantennae and IRAS 08572+3915 were observed
twice, and two flux measurements are shown separately.
Col.(2): Flux measurement in this work in mJy. 
Col.(3): Spitzer IRS flux at 18 $\mu$m.
References are \citet{rou06}, \citet{bra06}, \citet{bei06}, 
\citet{ima07a}, \citet{wu09}, and this paper (Figure 3).
Col.(4): Standard star used for flux calibration.
Col.(5): Adopted standard star flux at the observed wavelength in Jy.
}

\end{deluxetable}

%\clearpage

%%%%%%%%%% Table 6 %%%%%%%%%
\begin{deluxetable}{llcccc}
%\tabletypesize{\small}
\tabletypesize{\scriptsize}
\tablecaption{Emission size and surface brightness of LIRGs \label{tbl-6}}
\tablewidth{0pt}
\tablehead{
\colhead{Object} & \colhead{PSF standard} & \colhead{Size} &
\colhead{Physical size} & \colhead{Surface brightness} &
\colhead{$\tau_{18}$} \\  
\colhead{} & \colhead{} & \colhead{(arcsec)} & \colhead{(pc)} &
\colhead{(L$_{\odot}$ kpc$^{-2}$)} \\ 
\colhead{(1)} & \colhead{(2)} & \colhead{(3)} & \colhead{(4)} &
\colhead{(5)} & \colhead{(6)}  
}
\startdata 
Superantennae (1) & HR 7383 & $<$0.14 & $<$160 & $>$1.0$\times$10$^{13}$
& \nodata \\ 
IRAS 20551$-$4250 & HR 7869 & $<$0.15 & $<$120 &
$>$0.72$\times$10$^{13}$ $\Rightarrow$ $>$1.6$\times$10$^{13}$ & 0.7 \\ 
IRAS 01004$-$2237 & HR 400  & $<$0.13 & $<$255 &
$>$0.89$\times$10$^{13}$ & \nodata \\ 
IRAS 04103$-$2838 & HR 1231 & $<$0.09 & $<$165 &
$>$1.8$\times$10$^{13}$ & \nodata \\ 
VV 114 E-1         & HR 334 & $<$0.19 & $<$70  &
$>$0.20$\times$10$^{13}$ & \nodata \\
VV 114 E-2         & HR 334 & 0.78 & 295 & $\sim$0.02$\times$10$^{13}$ &
\nodata \\ 
NGC 1377          & HR 1231 & $<$0.14 & $<$16  &
$>$0.92$\times$10$^{13}$ $\Rightarrow$ $>$2.5$\times$10$^{13}$ & 0.85 \\ 
Superantennae (2) & HR 7383 & $<$0.19 & $<$210 &
$>$0.67$\times$10$^{13}$ & \nodata \\  
IRAS 00091$-$0736 & HR 117 & $<$0.18 & $<$350 & $>$0.07$\times$10$^{13}$
$\Rightarrow$ $>$0.24$\times$10$^{13}$ & 1.05 \\  
IRAS 00188$-$0856 & HR 117 & $<$0.10 & $<$215 & $>$0.39$\times$10$^{13}$
$\Rightarrow$ $>$0.75$\times$10$^{13}$ & 0.55 \\  
IRAS 01298$-$0744 & HR 334 & $<$0.18 & $<$400 & $>$0.08$\times$10$^{13}$
$\Rightarrow$ $>$0.30$\times$10$^{13}$ & 1.1 \\  
ESO 602$-$G025     & HR 8015 & 0.4 & 185 & $\sim$0.03$\times$10$^{13}$ &
\nodata \\  
NGC 5253  & HR 5287 & 0.32 & 6.3 & $\sim$1.5$\times$10$^{13}$ & \nodata \\  
NGC 7592 W & HR 8841 & $<$0.19 & $<$85 & $>$0.20$\times$10$^{13}$ & \nodata \\
\hline
IRAS 08572$+$3915 (1) & HR 3275 & $<$0.12 \tablenotemark{a} & $<$125 &
$>$1.1$\times$10$^{13}$ $\Rightarrow$ $>$3.3$\times$10$^{13}$ & 0.9 \\  
NGC 1614          & HR 1231 & 1.9 & 580 & $\sim$0.02$\times$10$^{13}$ &
\nodata \\  
NGC 2623          & HR 3275 & $<$0.34 & $<$115 &
$>$0.07$\times$10$^{13}$ & \nodata \\ 
IRAS 08572$+$3915 (2)   & HR 3705 & $<$0.18 \tablenotemark{a} & $<$190 &
$>$0.45$\times$10$^{13}$ $\Rightarrow$ $>$1.3$\times$10$^{13}$ & 0.9 \\   
IRAS 15250$+$3609   & HR 5763 & $<$0.05 & $<$50 & $>$3.4$\times$10$^{13}$
$\Rightarrow$ $>$7.0$\times$10$^{13}$ & 0.6 \\  
Arp 193   & HR 5219 & 1.0 & 420 & $\sim$0.009$\times$10$^{13}$ & \nodata
\\ \hline
\enddata

\tablenotetext{a}{\citet{soi00} constrained to be $<$0$\farcs$22.}

\tablecomments{
Col.(1): Object name.
The horizontal solid line separates sources observed with Gemini-South 
(above the line) and Subaru (below the line).
Col.(2): PSF standard star. 
Col.(3): Spatial extent of LIRG 18 $\mu$m emission in apparent size in
arcsec. 
Col.(4): Spatial extent of LIRG 18 $\mu$m emission in physical size in
pc. 
Col.(5): Emission surface brightness in L$_{\odot}$ kpc$^{-2}$ estimated
from the 18 $\mu$m flux (luminosity) and measured size for the 
18 $\mu$m emission regions. 
Values after correction for the 18 $\mu$m flux-attenuation are also
shown for LIRGs that display clearly detectable 18 $\mu$m silicate dust 
absorption features. 
No estimate is attempted for NGC 4945 because of its complex morphology.
Col.(6): Optical depth of the 18 $\mu$m silicate dust absorption
feature for detected sources, estimated based on our method (Imanishi et
al. 2007a; $\S$4.2 of this paper). 
}

\end{deluxetable}

%%%%%%%%%% Table 7 %%%%%%%%%
\begin{deluxetable}{lclcrrc}
\tabletypesize{\scriptsize}
\tablecaption{Flux, size, and emission surface brightness of sources
observed at $\sim$10 $\mu$m\label{tbl-7}}
\tablewidth{0pt}
\tablehead{
\colhead{Object} & \colhead{Flux (This work)} &
\colhead{Standard} & \colhead{Std flux} &
\colhead{Size} &
\colhead{Physical size} & \colhead{Surface brightness} \\
\colhead{} & \colhead{(mJy)} & \colhead{star} & \colhead{(Jy)} &
\colhead{(arcsec)} & \colhead{(pc)} & \colhead{(L$_{\odot}$ kpc$^{-2}$)} \\ 
\colhead{(1)} & \colhead{(2)} & \colhead{(3)} & \colhead{(4)} 
& \colhead{(5)} & \colhead{(6)} & \colhead{(7)} 
}
\startdata 
I Zw1    & 380  & HR 106 & 4.35 & $<$0.21 & $<$220 &
$>$0.52$\times$10$^{13}$ $\Rightarrow$ $>$2.6$\times$10$^{13}$ \\      
Mrk 1014 &  80  & HR 500 & 5.15 & $<$0.14 & $<$360 &
$>$0.33$\times$10$^{13}$ $\Rightarrow$ $>$1.6$\times$10$^{13}$ \\ 
PG 0844$+$349 & 50  & HR 3275 & 20.3 & $<$0.10 & $<$110 & $>$0.29$\times$10$^{13}$ $\Rightarrow$ $>$1.4$\times$10$^{13}$ \\  
PG 2130$+$099 & 135 & HR 8287 & 3.90 & $<$0.28 & $<$300 & $>$0.10$\times$10$^{13}$ $\Rightarrow$ $>$0.5$\times$10$^{13}$ \\  
PG 2214$+$139 & 75 & HR 8287 & 3.90 & $<$0.17 & $<$190 &
$>$0.16$\times$10$^{13}$ $\Rightarrow$ $>$0.8$\times$10$^{13}$ \\ 
Mrk 266 N & 92 & HR 5219 & 46.54 & 0.56 & 300 & 0.02$\times$10$^{13}$ $\Rightarrow$ 0.1$\times$10$^{13}$ \\ 
Mrk 266 S & 20 & HR 5219 & 46.54 & \nodata & \nodata & \nodata \\
\enddata

\tablecomments{
Col.(1): Object name.
Col.(2): Flux measurement at $\sim$10 $\mu$m ($N11.7$ or $N8.8$) in mJy. 
Col.(3): Standard star used for flux and PSF calibration.
Col.(4): Adopted standard star flux at the observed wavelength in Jy.
Col.(5): Spatial extent at $\sim$10 $\mu$m in apparent size in arcsec. 
Col.(6): Spatial extent at $\sim$10 $\mu$m in physical size in pc. 
Col.(7): Emission surface brightness in L$_{\odot}$ kpc$^{-2}$ estimated
from the observed 11.7 $\mu$m (or 8.8 $\mu$m) flux (luminosity) and 
the physical size of emission regions.
Values after correction for the infrared-unprobed emission (= fraction 
of energetic radiation unabsorbed by dust) are also added for sources 
classified optically as Seyferts ($\S$5.1.4).
Mrk 266 S is barely detected at 10 $\mu$m, but is too faint to estimate
the spatial extent and emission surface brightness in a reliable manner.
}
\end{deluxetable}

\clearpage

%---  Figure 1 ---%
\begin{figure}
\includegraphics[angle=0,scale=.6]{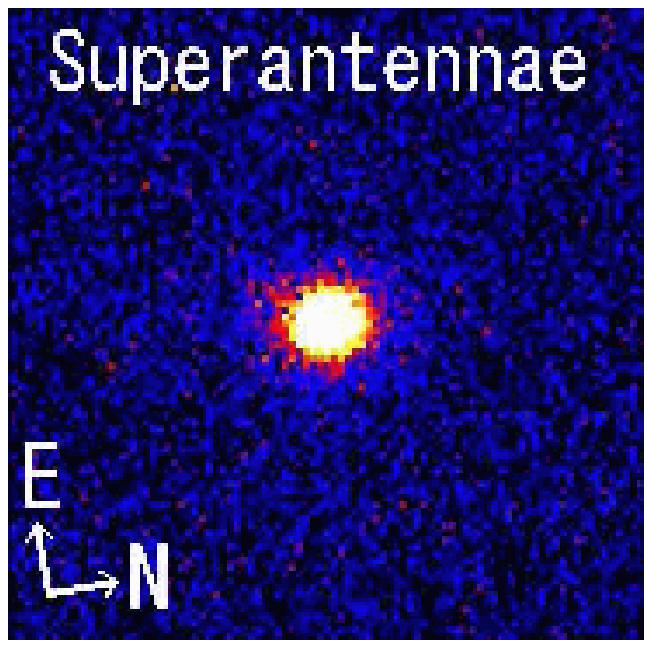} \hspace{0.1cm} 
\includegraphics[angle=0,scale=.6]{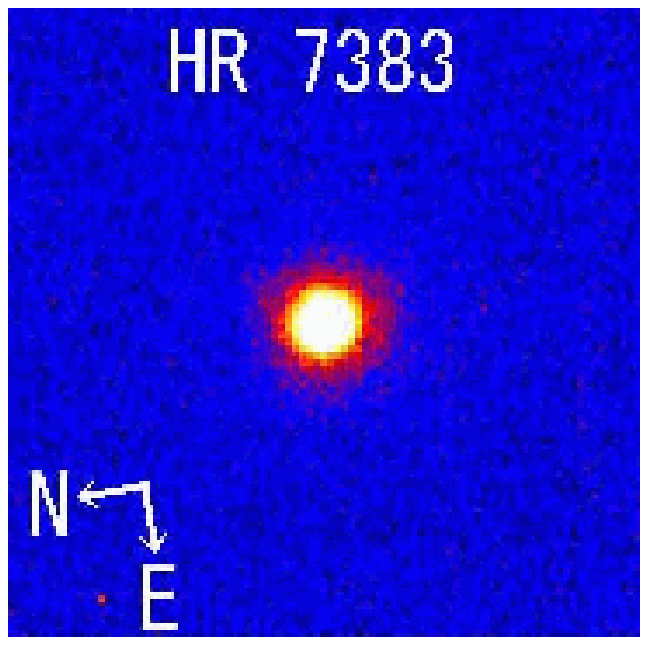}
\hspace{0.1cm}
\includegraphics[angle=0,scale=.6]{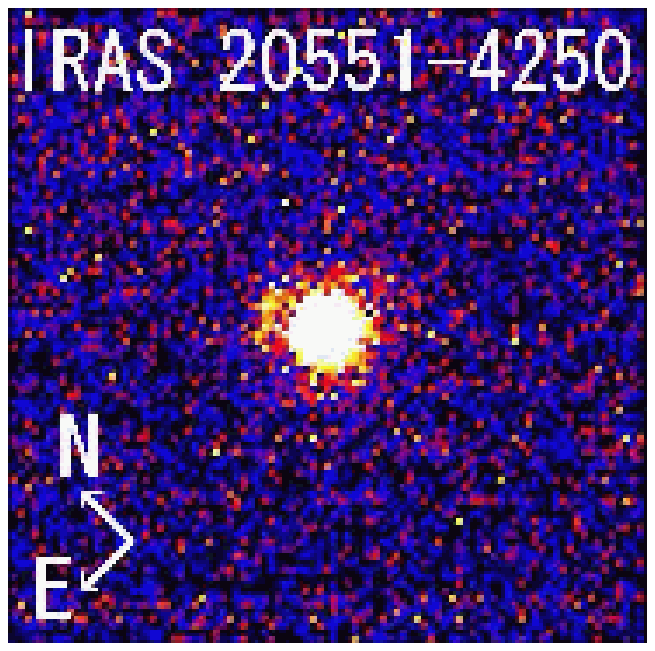}
\hspace{0.1cm} 
\includegraphics[angle=0,scale=.6]{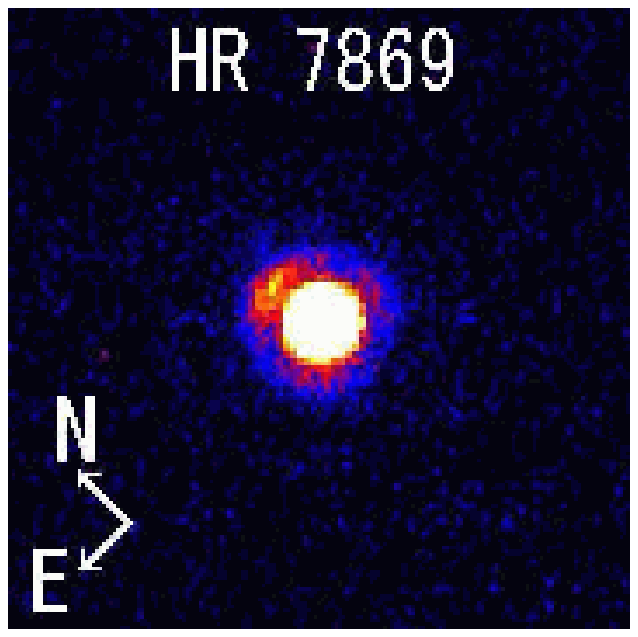} \vspace*{0.3cm}\\ 
\includegraphics[angle=0,scale=.6]{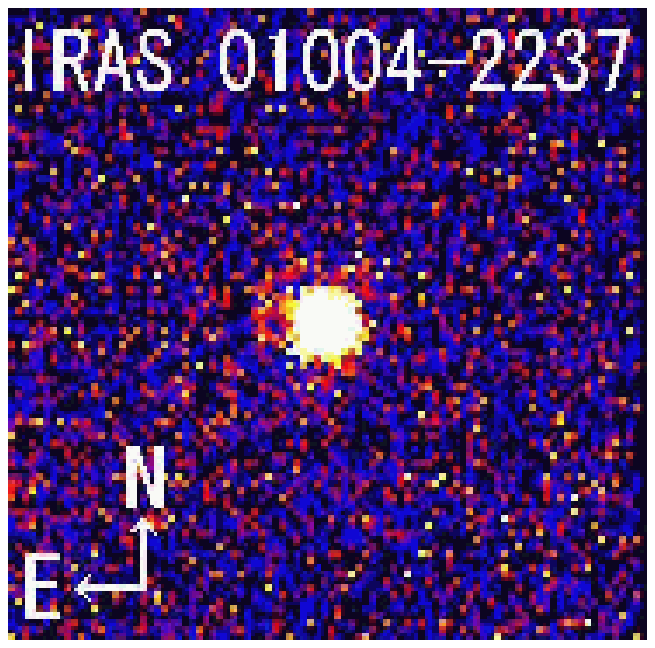}
\hspace{0.1cm}
\includegraphics[angle=0,scale=.6]{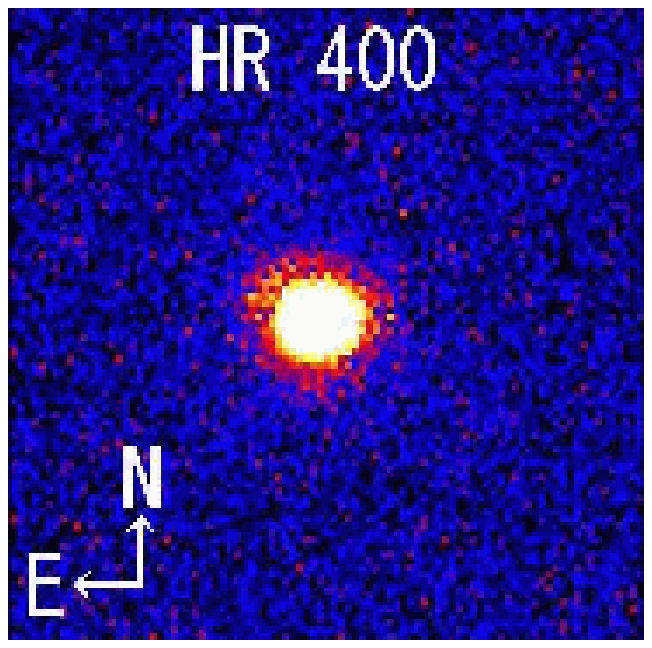} 
\hspace{0.1cm} 
\includegraphics[angle=0,scale=.6]{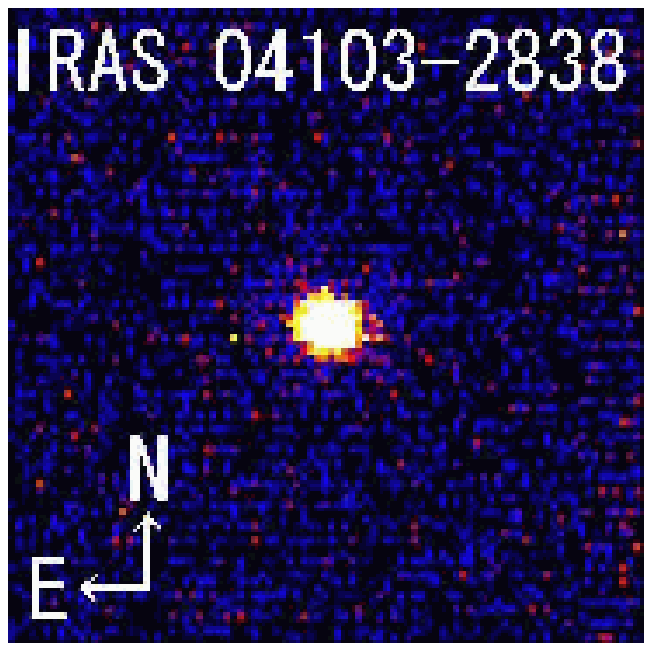} 
\hspace{0.1cm}
\includegraphics[angle=0,scale=.6]{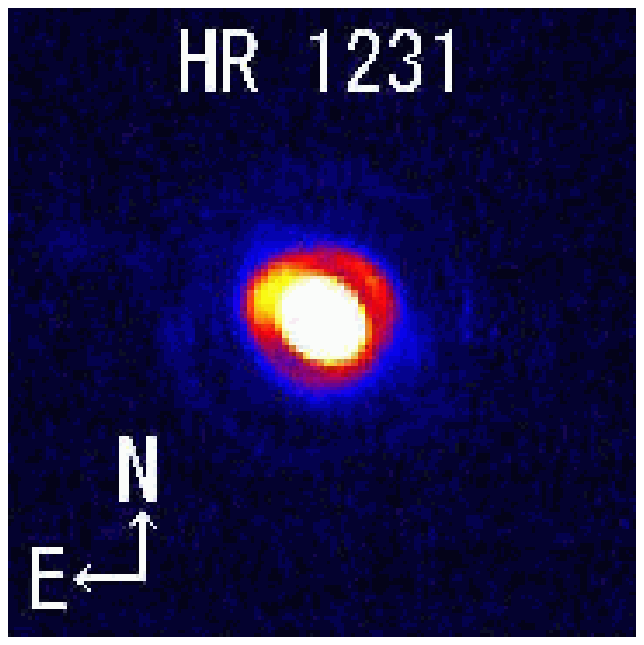}\vspace*{0.3cm} \\
\includegraphics[angle=0,scale=.6]{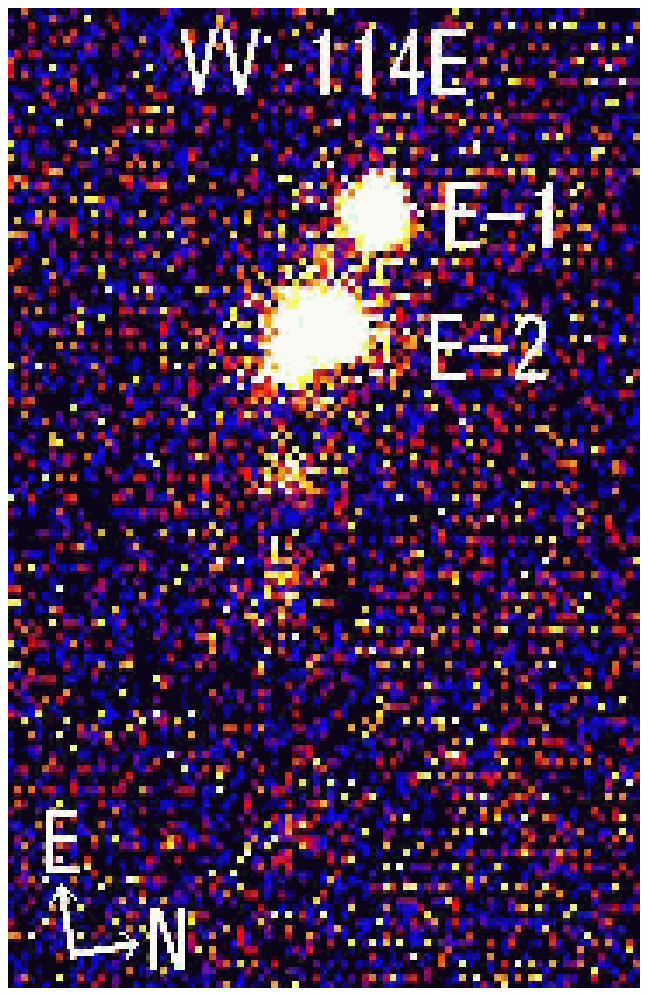}
\hspace{0.1cm} 
\includegraphics[angle=0,scale=.6]{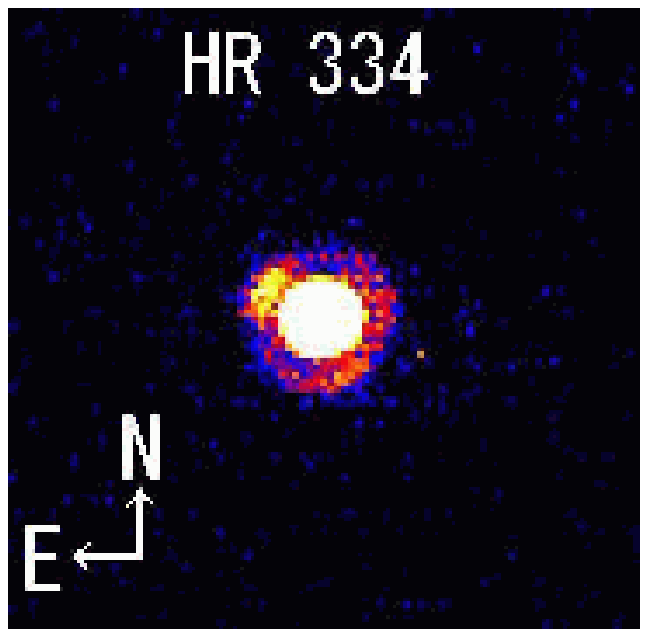}\vspace{0.1cm}
\includegraphics[angle=0,scale=.6]{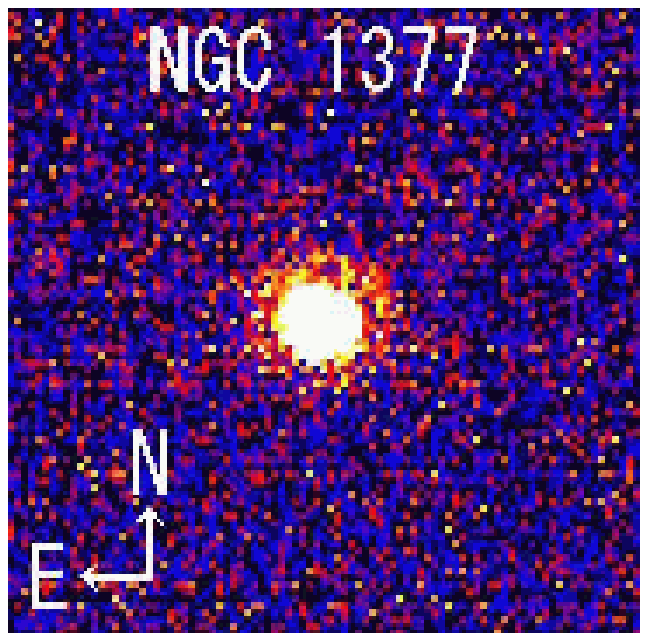}
\hspace{0.1cm}
\includegraphics[angle=0,scale=.6]{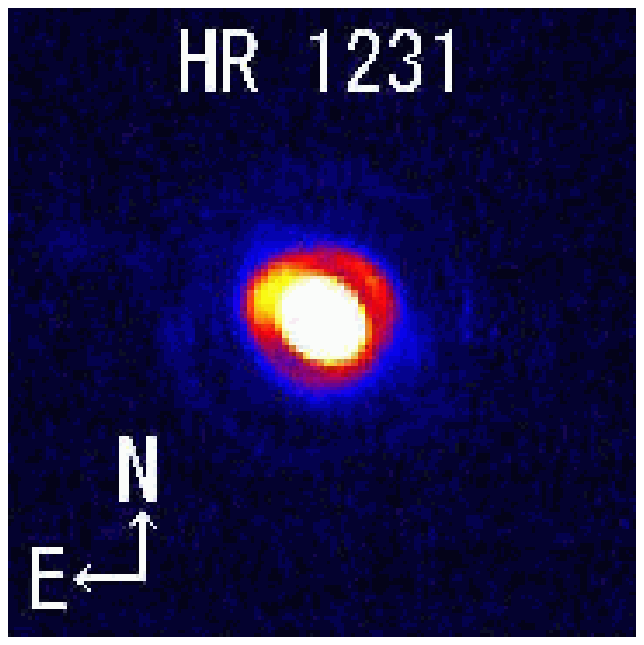} 
\hspace*{0.3cm} \\
\includegraphics[angle=0,scale=.6]{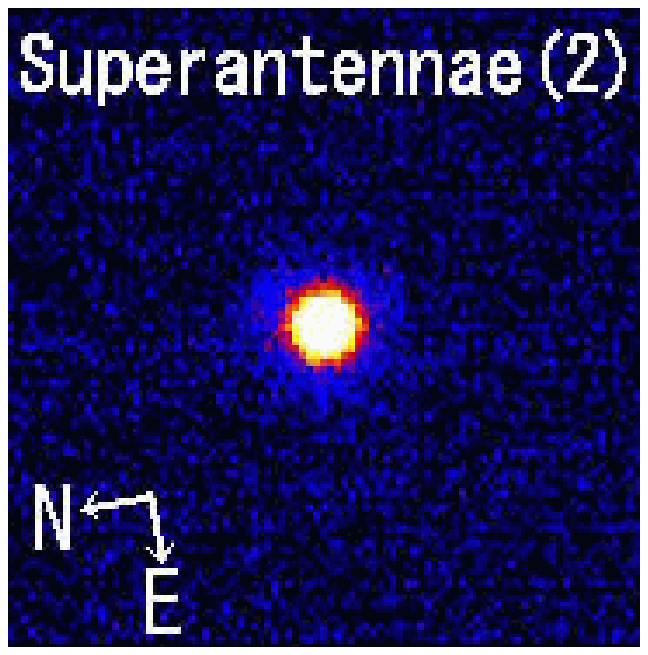} \hspace{0.1cm}
\includegraphics[angle=0,scale=.6]{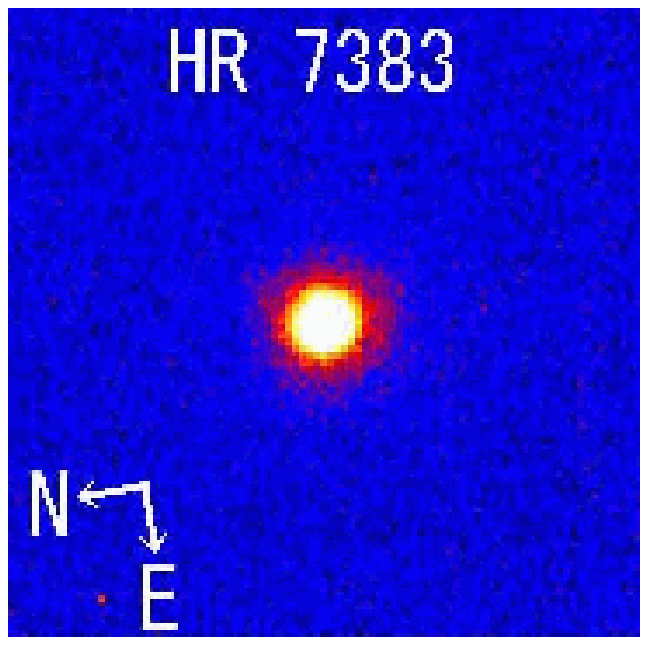} 
\hspace{0.1cm} 
\includegraphics[angle=0,scale=.6]{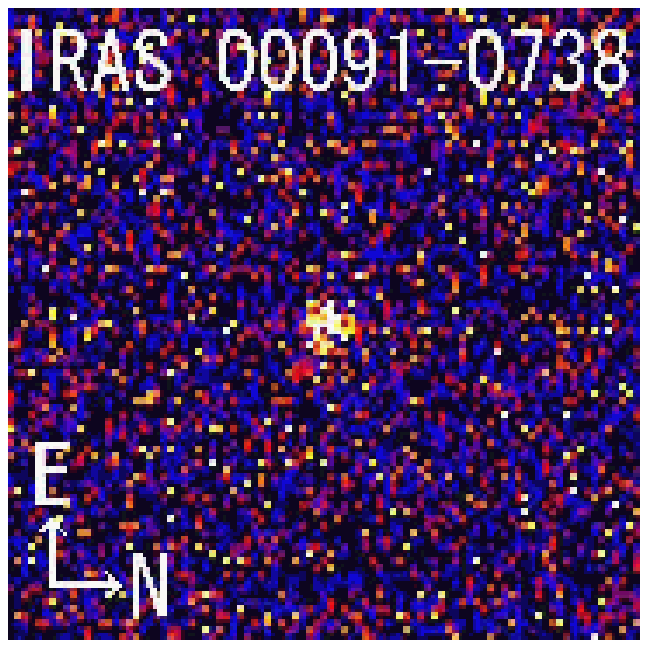}
\hspace{0.1cm} 
\includegraphics[angle=0,scale=.6]{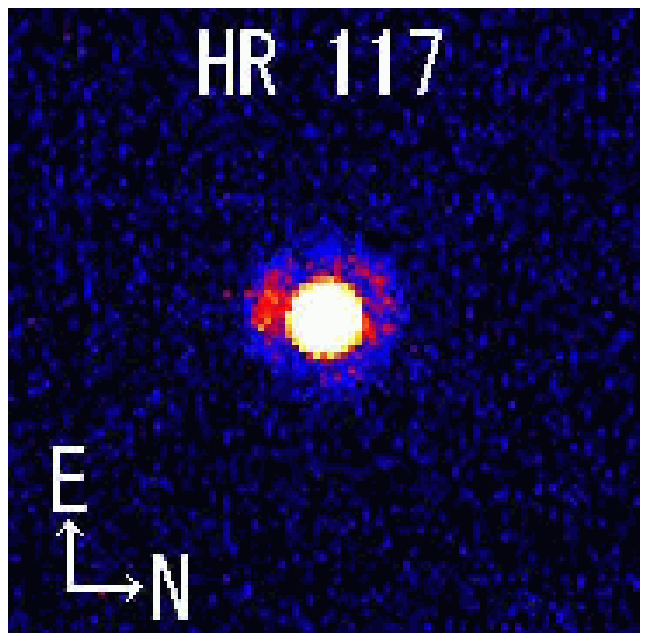} 
\end{figure}

\clearpage

\begin{figure}
\includegraphics[angle=0,scale=.6]{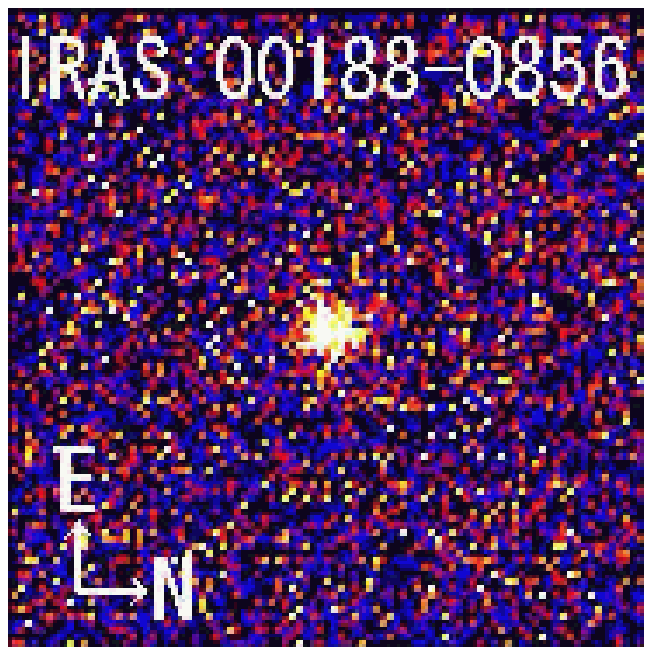}  
\hspace{0.1cm} 
\includegraphics[angle=0,scale=.6]{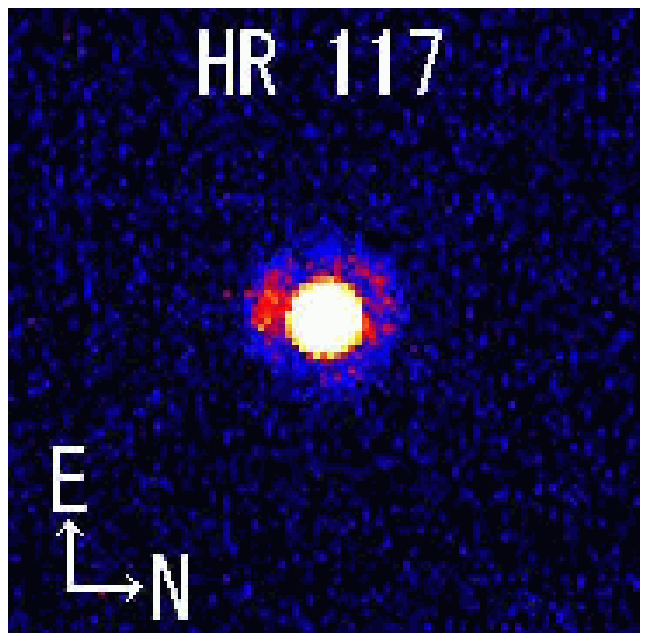} 
\hspace{0.1cm} 
\includegraphics[angle=0,scale=.6]{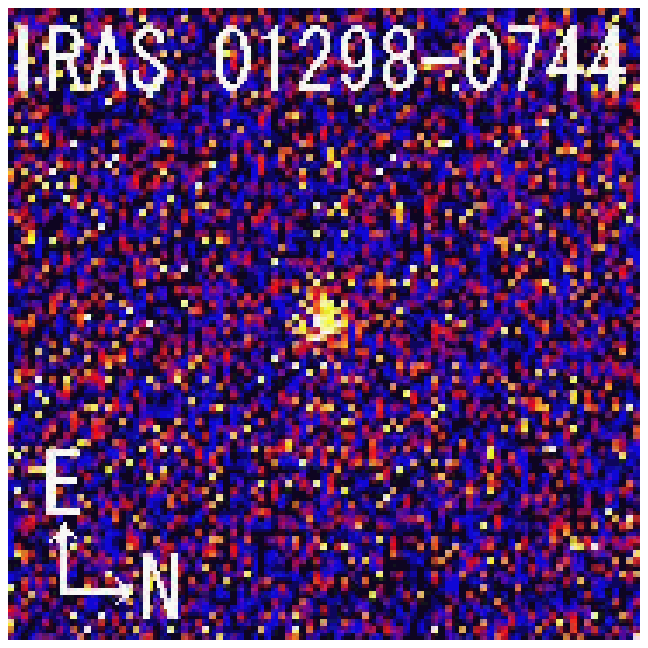}
\hspace{0.1cm} 
\includegraphics[angle=0,scale=.6]{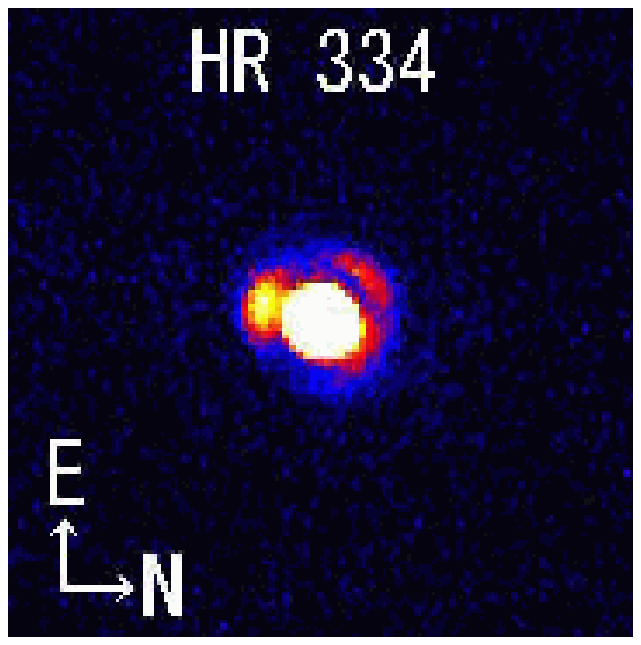} \vspace*{0.1cm}\\
\includegraphics[angle=0,scale=.6]{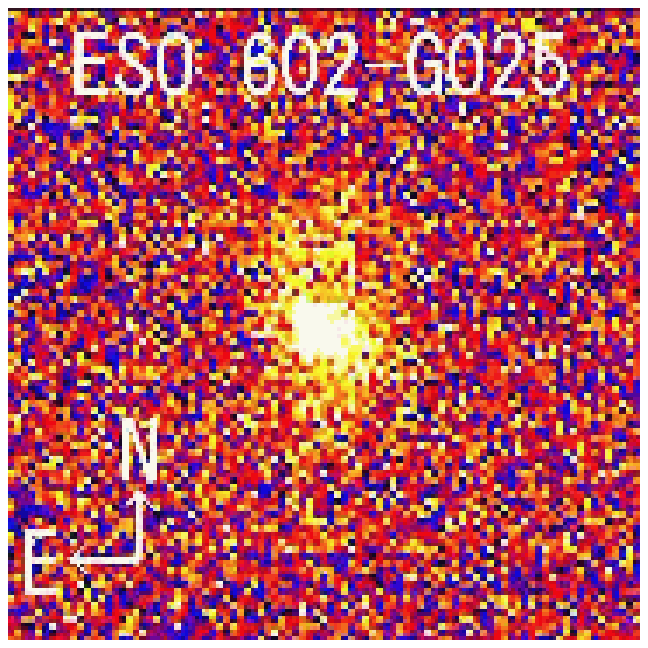} 
\hspace{0.1cm} 
\includegraphics[angle=0,scale=.6]{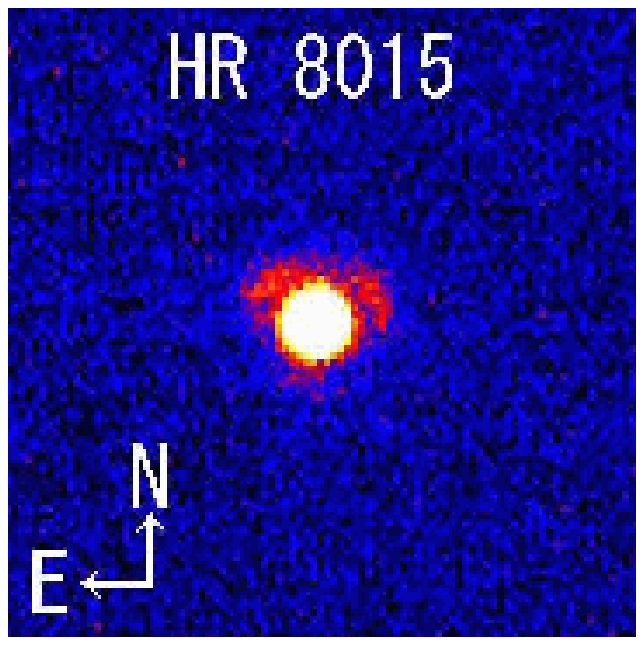} 
\hspace{0.1cm} 
\includegraphics[angle=0,scale=.6]{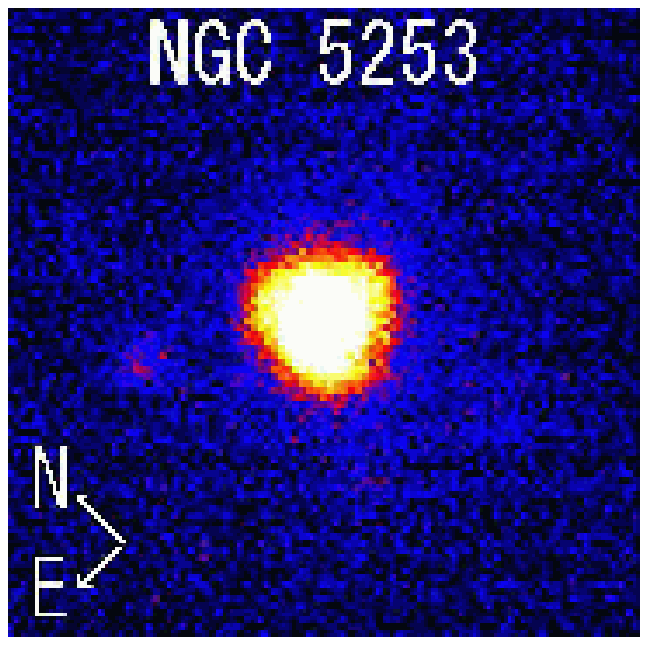} 
\hspace{0.1cm} 
\includegraphics[angle=0,scale=.6]{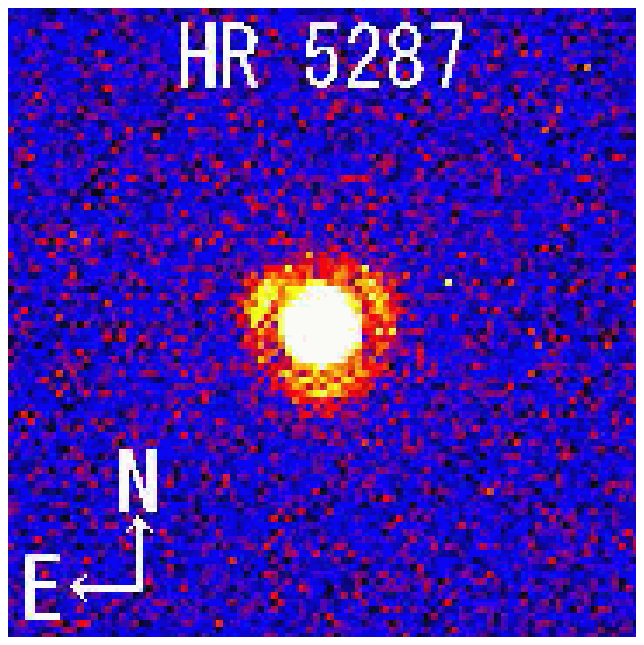} 
\vspace*{0.1cm} \\ 
\includegraphics[angle=0,scale=.6]{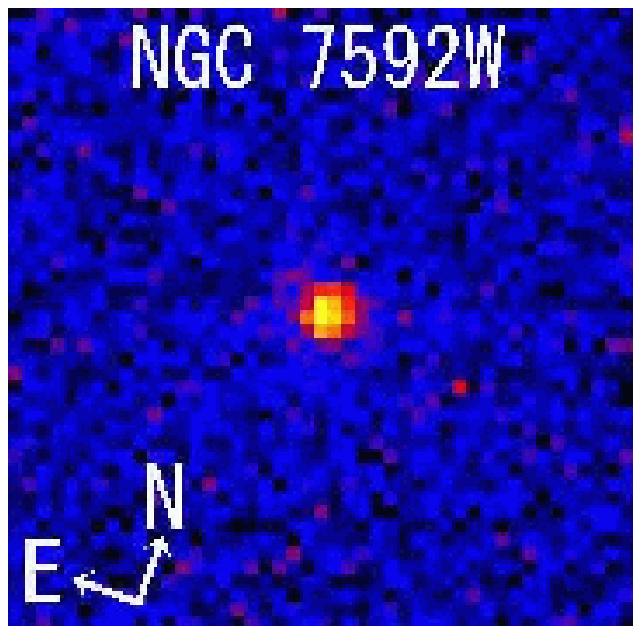}
\hspace{0.1cm} 
\includegraphics[angle=0,scale=.6]{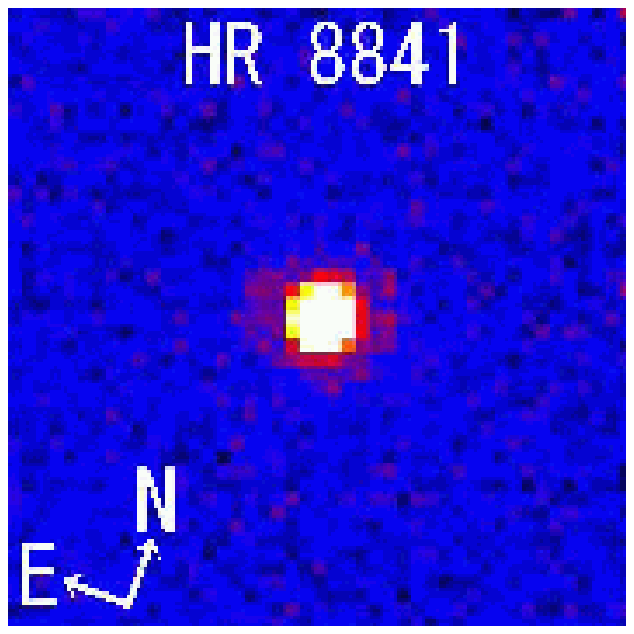}
\vspace*{0.1cm} \\
\includegraphics[angle=0,scale=.6]{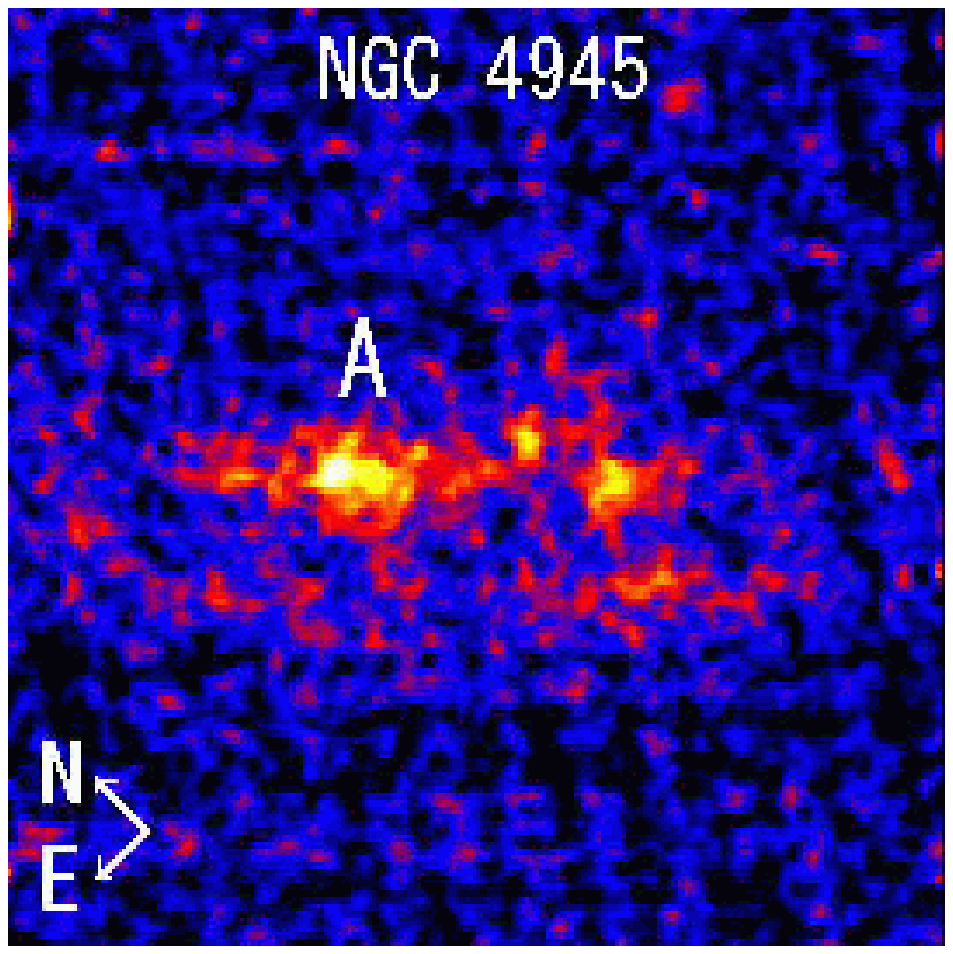}
\hspace{0.1cm} 
\includegraphics[angle=0,scale=.6]{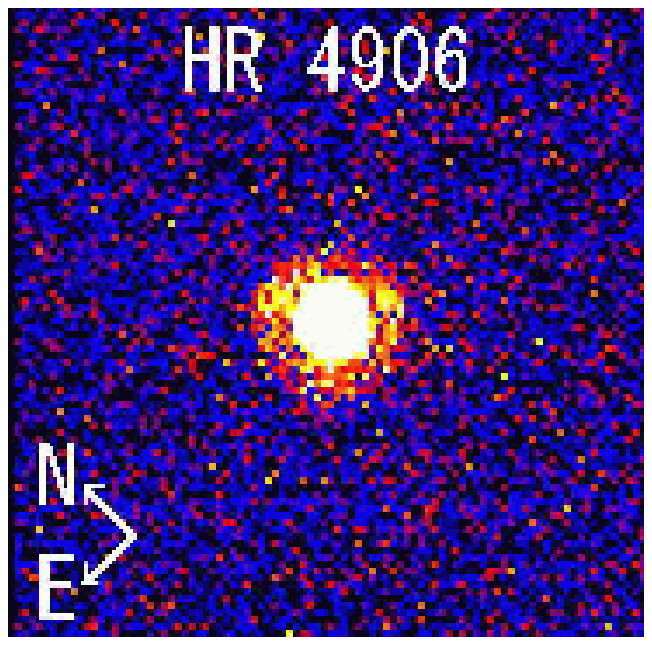}
\caption{
$Qa$ (18.3 $\mu$m) images of LIRGs and corresponding standard stars, 
taken with Gemini-South T-ReCS. 
The image size is 8'' $\times$ 8'', except VV 114 E and NGC
4945 (two larger figures) whose image sizes are 8'' $\times$ 12'' and
12'' $\times$ 12'', respectively, to include spatially extended
structures. 
The size of each image is proportional to the actual field-of-view.
The north (N) and east (E) directions are indicated. 
Upper and lower display levels are set arbitrarily and adjusted to
individual sources, to make interesting emission structures (e.g.,
diffraction ring pattern) visually clear. 
The 18 $\mu$m emission of NGC 4945 is spatially extended. 
To better visualize the morphology, the image is Gaussian-smoothed by 2
pixels $\times$ 2 pixels.
}
\end{figure}

%\clearpage

%---  Figure 2 ---%
\begin{figure}
\includegraphics[angle=0,scale=.44]{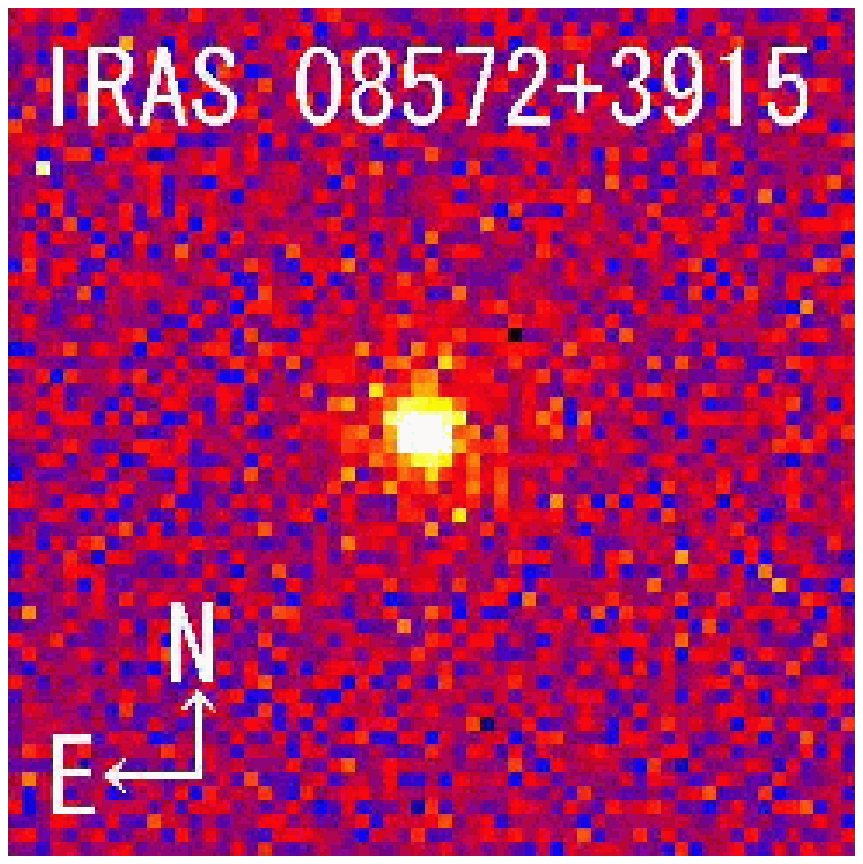}
\hspace{0.1cm}
\includegraphics[angle=0,scale=.44]{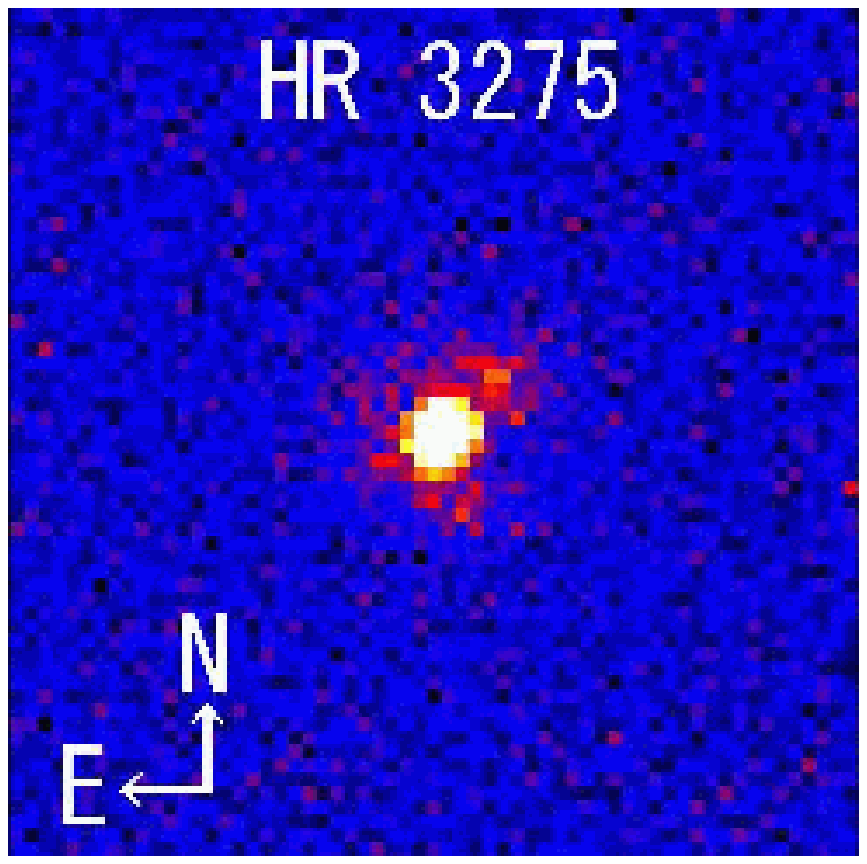}
\hspace{0.1cm}
\includegraphics[angle=0,scale=.44]{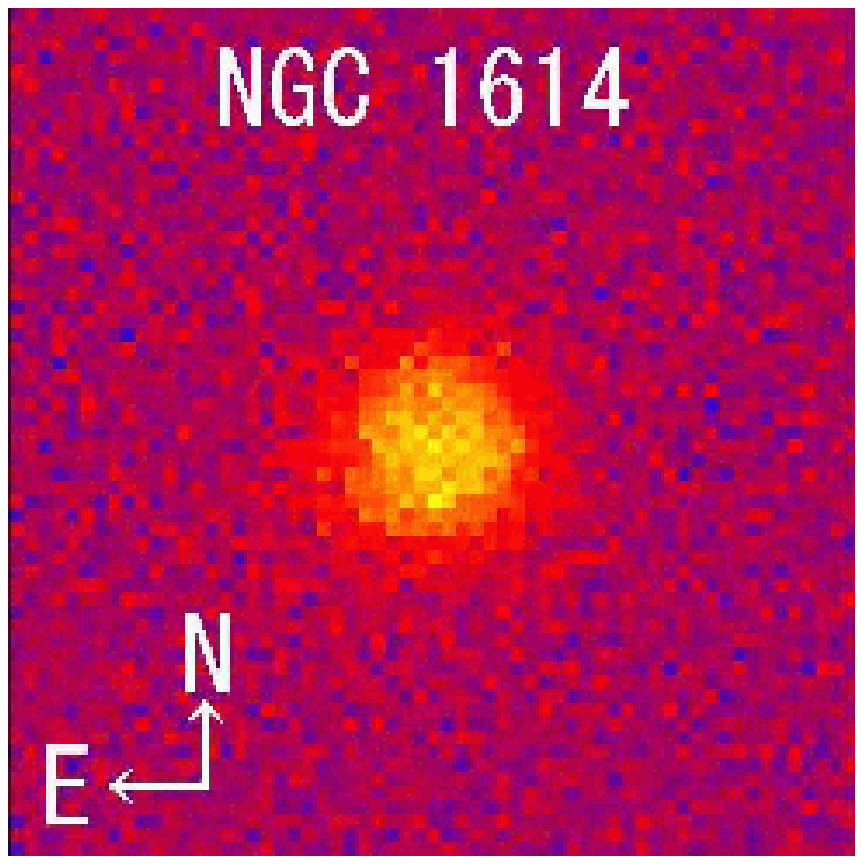}
\hspace{0.1cm}
\includegraphics[angle=0,scale=.44]{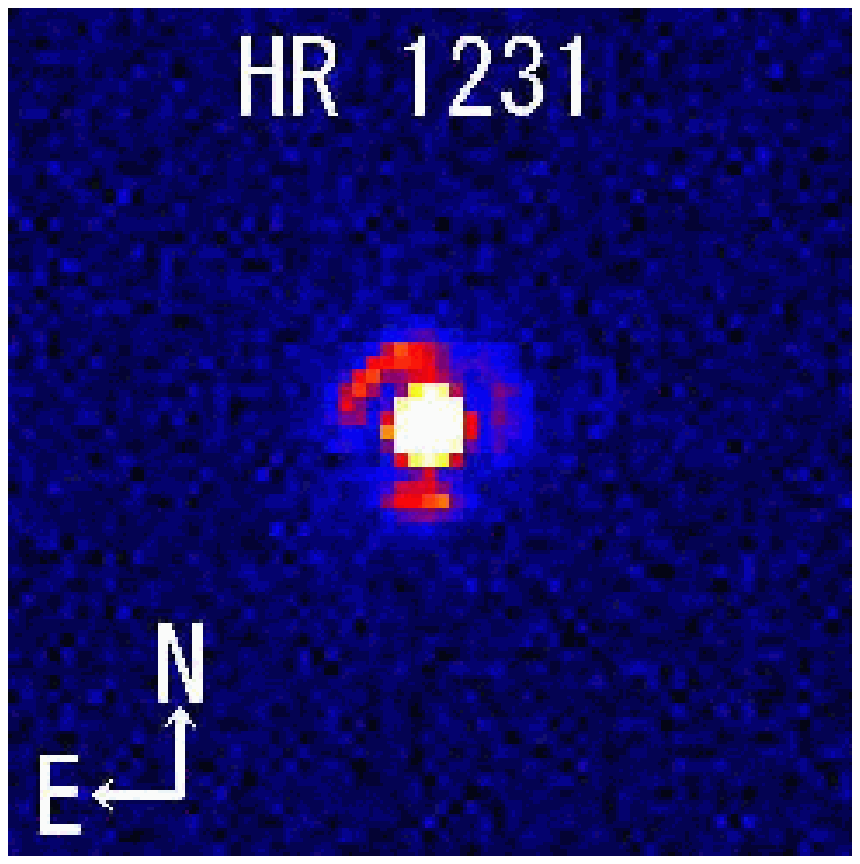} 
\vspace*{0.3cm} \\ 
\includegraphics[angle=0,scale=.44]{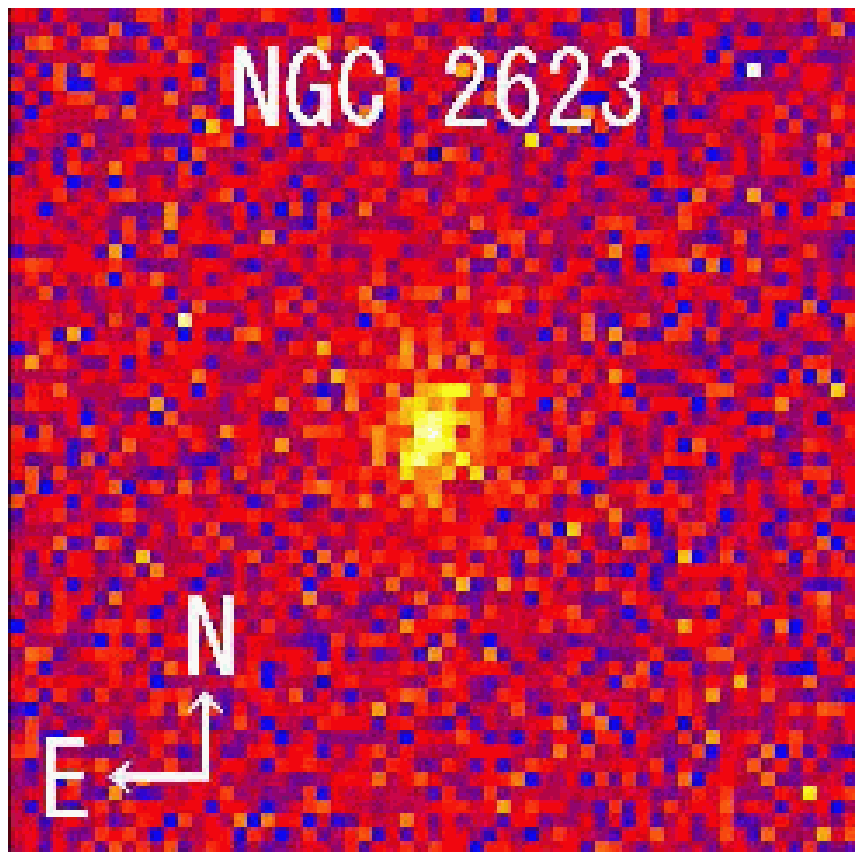}
\hspace{0.1cm} 
\includegraphics[angle=0,scale=.44]{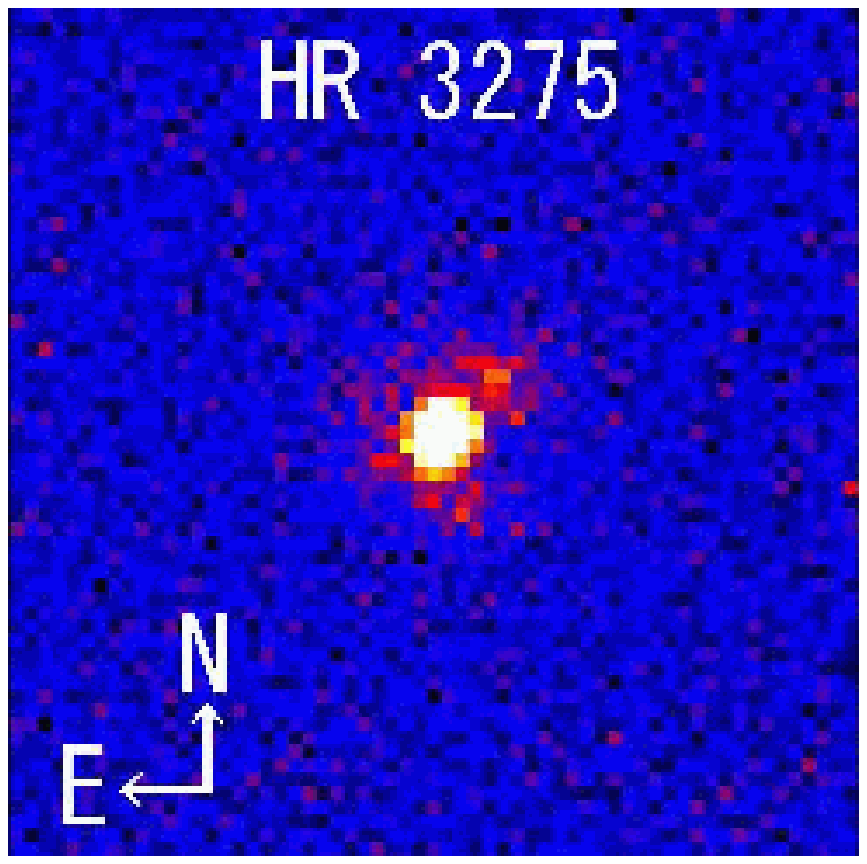} \vspace*{0.3cm}\\ 
\includegraphics[angle=0,scale=.44]{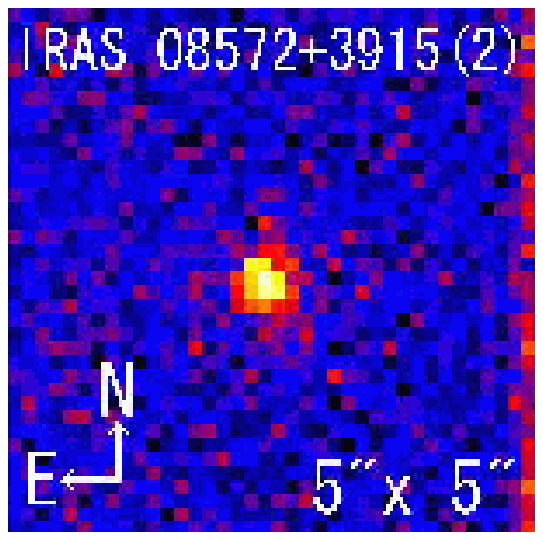}
\hspace{0.1cm}
\includegraphics[angle=0,scale=.44]{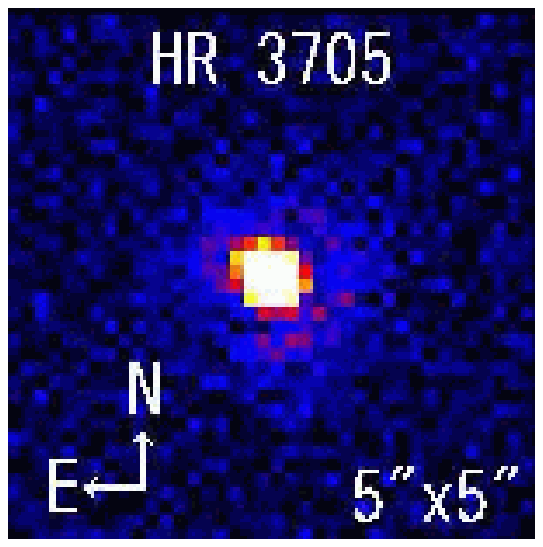}
\hspace{0.1cm}
\includegraphics[angle=0,scale=.44]{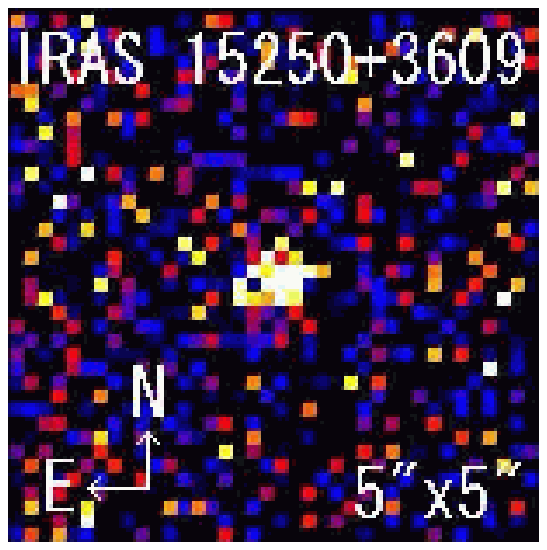}
\hspace{0.1cm}
\includegraphics[angle=0,scale=.44]{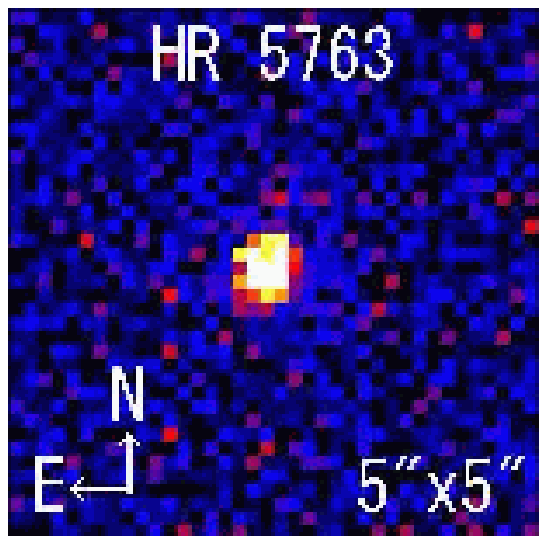} 
\hspace{0.1cm}
\includegraphics[angle=0,scale=.44]{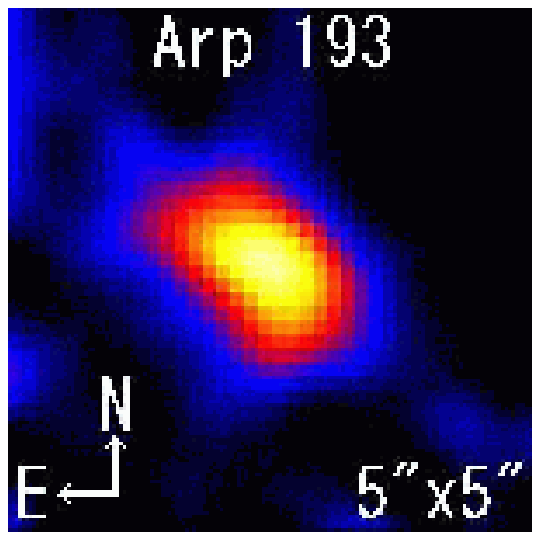}
\hspace{0.1cm}
\includegraphics[angle=0,scale=.44]{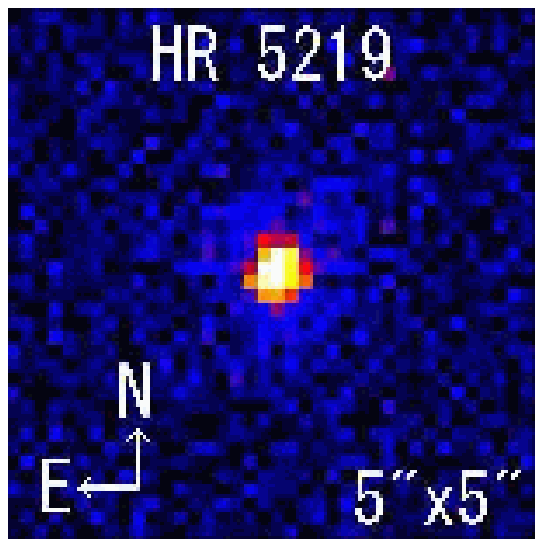}
\caption{
$Q$17.7 (17.7 $\mu$m) images of LIRGs and corresponding standard stars, 
obtained with Subaru COMICS. 
The image size is 8'' $\times$ 8'' for data taken in 2008, while it is
5'' $\times$ 5'' for data taken in 2009 (smaller figures in the bottom),
because only 50 rows were read out, due to high background
emission from Earth's atmosphere. 
The size of each image is proportional to the actual field-of-view.
The upper and lower display levels are varied, according to individual
sources, to show interesting emission patterns.
The 18 $\mu$m emission of Arp 193 is spatially extended.
To reduce noise and better visualize the morphology, the image is
Gaussian-smoothed by 2 pixels $\times$ 2 pixels.
}

\end{figure}

%--------------  Figure 3 -----------------%
\begin{figure}
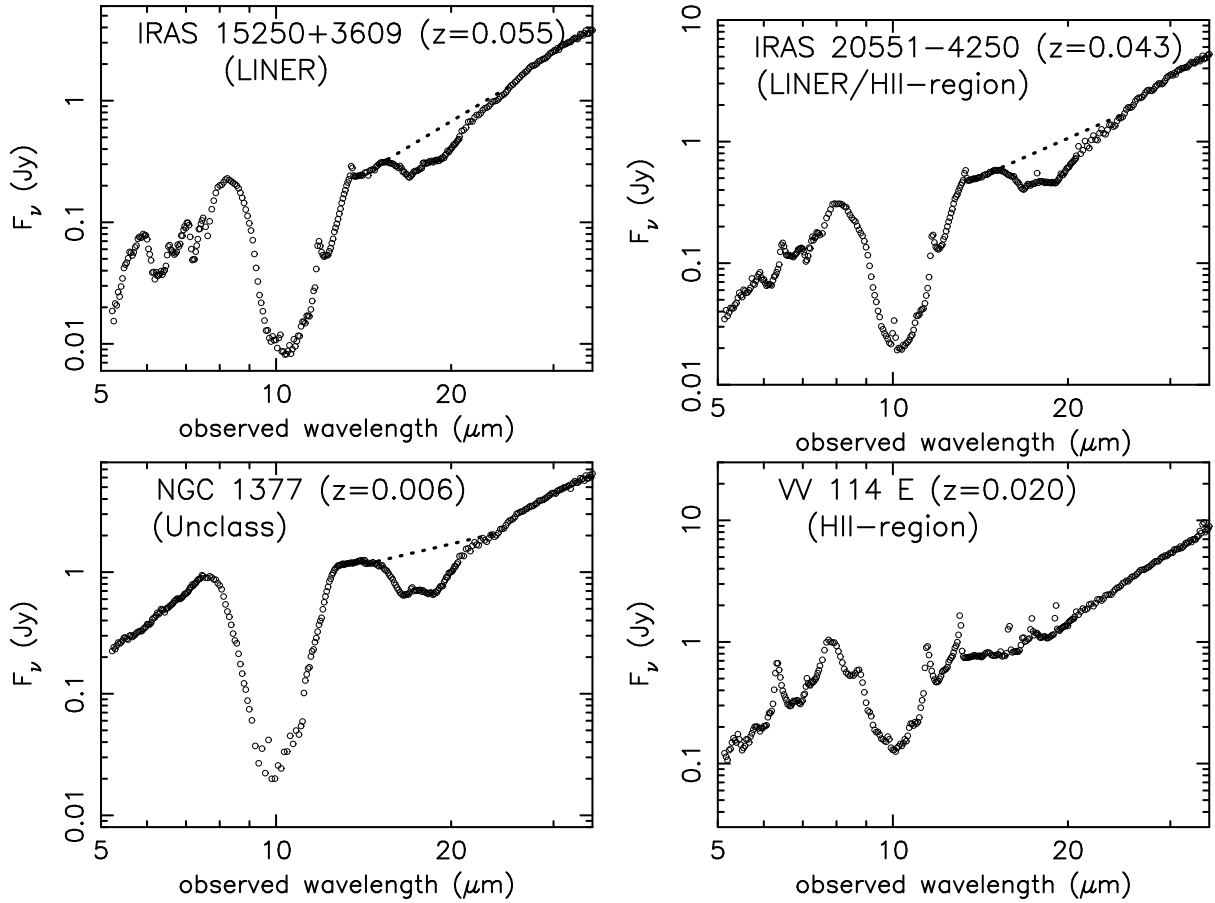

\includegraphics[angle=-90,scale=.35]{f3a.eps} \hspace{0.2cm} 
\includegraphics[angle=-90,scale=.35]{f3b.eps} \\ 
\includegraphics[angle=-90,scale=.35]{f3c.eps} \hspace{0.2cm} 
\includegraphics[angle=-90,scale=.35]{f3d.eps} \\
\caption{
Spitzer IRS low-resolution spectra of IRAS 15250+3609, IRAS
20551$-$4250, NGC 1377, and VV 114 E. 
The dotted line represents the power-law continuum to measure the
optical depths of the 18 $\mu$m silicate dust absorption features,
determined from data points at $\lambda_{\rm rest}$ = 14.2 $\mu$m and 24
$\mu$m \citep{ima07a,ima09,ima10a}.
}
\end{figure} 

\end{document}